\documentclass[hyper,letterpaper]{JHEP3}
\usepackage{amsmath,amssymb,amsfonts,bm}
\usepackage{graphicx}
\usepackage{multirow}
\usepackage{verbatim}
\usepackage{appendix}
\usepackage{rotating}
\usepackage{multirow}

\numberwithin{equation}{section}

\newcommand{\bea}{\begin{eqnarray}}
\newcommand{\eea}{\end{eqnarray}}
\newcommand{\be}{\begin{equation}}
\newcommand{\ee}{\end{equation}}

\newcommand{\Z}{{\mathbb Z}}
\newcommand{\R}{{\mathbb R}}
\newcommand{\C}{{\mathbb C}}

\newcommand{\Tr}{{\rm Tr \,}}

\newcommand{\CA}{\mathcal{A}}
\newcommand{\CB}{\mathcal{B}}
\newcommand{\CC}{\mathcal{C}}

\newcommand{\CE}{\mathcal{E}}
\newcommand{\CF}{\mathcal{F}}
\newcommand{\CG}{\mathcal{G}}
\newcommand{\CH}{\mathcal{H}}
\newcommand{\CI}{\mathcal{I}}

\newcommand{\CK}{\mathcal{K}}
\newcommand{\CL}{\mathcal{L}}
\newcommand{\CM}{\mathcal{M}}
\newcommand{\CN}{\mathcal{N}}
\newcommand{\CO}{\mathcal{O}}

\newcommand{\CQ}{\mathcal{Q}}

\newcommand{\CS}{\mathcal{S}}
\newcommand{\CT}{\mathcal{T}}

\newcommand{\CW}{\mathcal{W}}

\newcommand{\CZ}{\mathcal{Z}}

\newcommand{\cp}{{\mathbb{C}}{\mathbf{P}}}
\def\TT{{\Bbb{T}}}
\def\LL{{\Bbb{L}}}
\def\Weyl{{\mathcal W}}

\def\btimes{~{{{\lower1pt\hbox{$\square$}} \kern-7.6pt \times}}~}

\newcommand{\ttL}{{\mathtt L}}
\newcommand{\ttR}{{\mathtt R}}
\newcommand{\fq}{{\mathfrak q}}
\newcommand{\fp}{{\mathfrak p}}
\newcommand{\ft}{{\mathfrak t}}

\title{Surface Operators}

\author{Sergei Gukov$^{1,2}$
\\ ~
\\
$^1$ California Institute of Technology, Pasadena, CA 91125, USA \\
$^2$ Walter Burke Institute for Theoretical Physics, California Institute of Technology, Pasadena, CA 91125
}

\abstract{We give an introduction and a broad survey of surface operators in 4d gauge theories,
with a particular emphasis on aspects relevant to AGT correspondence.
One of the main goals is to highlight the boundary between what we know and what we don't know about surface operators.
To this end, the survey contains many open questions and suggests various directions for future research.
Although this article is mostly a review, we did include a number of new results, previously unpublished.
\\
\\
\\
\\
\\
\\
{\tt CALT-TH-2014-170}
}

\begin{document}


\section{What is a surface operator?}
\label{sec:what}

Surface operators (a.k.a. surface defects) in a four-dimensional gauge theory
are operators supported on two-dimensional submanifolds in the space-time manifold $M$.
They are particular examples of non-local operators in quantum field theory (QFT) that
play the role of ``thermometers'' in a sense that, when introduced in the Feynman path integral,
their correlation functions provide us with valuable information about the physics
of a QFT in question (phases, non-perturbative phenomena, {\it etc.}).

In general, non-local operators can be classified by dimension (or, equivalently, codimension)
of their support, which in four dimensions clearly can range from zero to four, so that we have
the following types of operators:
\begin{itemize}
\item codimension 4: the operators of codimension 4 are the usual
local operators $\CO (p)$ supported at a point $p \in M$.
These are the most familiar operators in this list,
which have been extensively studied {\it e.g.} in the context of
the AdS/CFT correspondence. Typical examples of local operators
can be obtained by considering gauge-invariant combinations of
the fields in the theory, {\it e.g.} $\CO (p) = \Tr (\phi^n \ldots)$.
\item codimension 3: line operators. Important examples of
such operators are Wilson and 't Hooft operators,
which are labeled, respectively, by a representation, $R$, of
the gauge group, $G$, and by a representation $^LR$ of the dual
gauge group $^LG$.
\item codimension 2: surface operators. These are perhaps
least studied among the operators and defects listed here,
and will be precisely our main subject.
\item codimension 1: domain walls and boundaries.
\end{itemize}
\noindent
After giving the reader a basic idea about different types of non-local operators
classified by (co)dimension of their support, perhaps it is worth mentioning that
some of them --- usually called ``electric'' --- can be constructed directly from
elementary fields present in the path integral formulation of the theory.
In the above classification, we already mentioned examples of such operators that are actually local,
{\it i.e.} any gauge-invariant combination of elementary fields gives an example.
Among non-local operators, a typical example of ``electric'' operators
is a Wilson line operator labeled by a representation $R$ of the gauge group $G$:
\begin{equation}
W_R (K) = \Tr_R ~{\rm Hol}_{K} (A)
= \Tr_R \left( P \! \exp \oint A \right)
\label{wilson}
\end{equation}
Another type of operators, called ``magnetic'' (a.k.a. disorder operators) can not be defined via
(algebraic) combinations of elementary fields and calls for alternative definitions,
which will be considered below and which will be crucial for defining surface operators.

A surface operator in four-dimensional gauge theory is
an operator supported on a 2-dimensional submanifold $D \subset M$
in the space-time manifold $M$.
In other words, according to the above classification,
it is an operator whose dimension and codimension are both equal to 2:
\begin{equation}
4 = 2 + 2
\label{224}
\end{equation}
This simple equation illustrates how the dimension of the space-time manifold $M$
splits into the tangent and normal spaces to the support, $D$, of the surface operator.
Note, that 2 also happens to be the degree of the differential form $F$,
the curvature of the gauge field $A$.
This basic fact and equation \eqref{224} make surface operators somewhat special
in the context of 4d gauge theory.

Indeed, since the degree of the 2-form $F$ matches the dimension of the tangent as well as normal space to $D \subset M$,
we can either write an integral
\begin{equation}
\exp \left( i\eta \int_D F \right)
\label{etafirst}
\end{equation}
which defines an electric surface operator analogous to \eqref{wilson} in abelian $U(1)$ gauge theory,
or write a relation
\begin{equation}
F = 2\pi \alpha \delta_D + \ldots
\label{alphafirst}
\end{equation}
where $\delta_D$ is a 2-form delta-function Poincar\'e dual to $D$.
In \eqref{etafirst} and \eqref{alphafirst} we used the basic fact that, respectively,
dimension and codimension of the surface operator (or, to be more precise, its support) equals the degree of the differential form $F$.
These relations define magnetic (resp. electric) surface operators in abelian 4d gauge theory --- with any amount of supersymmetry, including $\CN=2$ that will be of our prime interest in this note ---
and admit a simple generalization to non-abelian theories that will be discussed shortly.

Already at this stage, however, it is a good idea to pause and ask the following questions that shall guide us in the exploration of surface defects:
\begin{itemize}
\item How can one define surface operators?
\item What are they classified by?
\item Are there supersymmetric surface operators?
\item What are the correlation functions of surface operators?
\item What is the OPE algebra of line operators in the presence of a surface operator?
\item How do surface operators transform under dualities?
\end{itemize}
The answer to many of these questions is not known at present, except in some special cases.
One such special case is that of abelian gauge theory with gauge group $G \cong U(1)^r = \LL$, where all of the above questions can be answered:
\begin{itemize}
\item By combining the above constructions \eqref{etafirst} and \eqref{alphafirst} for each $U(1)$ factor in $G$
one can produce a surface operator that, in general, preserves some part of the gauge group, $\LL \subseteq G$.
\item The resulting surface operators are labeled by a {\it discrete} choice of $\LL \subseteq G$
and two sets of {\it continuous} parameters
\be
(\alpha,\eta) \; \in \; \TT \times \TT^{\vee}
\label{aetaabelian}
\ee
where $\TT = G / \LL$ and $\TT^{\vee}$ is its dual.
\item They are compatible with any amount of supersymmetry and define half-BPS surface operators in SUSY gauge theories.
\item Physically, the world-volume $D$ of such a surface operator can be interpreted as a ``visible'' Dirac string
for a dyon with electric and magnetic charges $(\eta,\alpha)$ that do not obey Dirac quantization condition.
\item A remarkable property of abelian 4d gauge theory is that it enjoys electric-magnetic duality,
even in the absence of supersymmetry \cite{Witten:1995gf,Verlinde:1995mz}. This duality exchanges the role of $\alpha$ and $\eta$:
\be
(\alpha, \eta) \; \to \; (\eta, - \alpha)
\label{emduality}
\ee
\item A novel feature of surface operators is that they are labeled not only by discrete but also by continuous parameters.
A kink-like configuration within the surface operator that represents an adiabatic change of continuous parameters
along a closed loop in the parameter space \eqref{aetaabelian} represents a Wilson-'t Hooft line operator.
In other words, line operators correspond to closed loops in the space of continuous parameters and are labeled by elements of
\begin{equation}
\pi_1 \big( \{ {\rm parameters} \} \big)
\label{fundparam}
\end{equation}
\end{itemize}
Many of these aspects have analogues in non-abelian gauge theory, where essential features may look similar
though dressed with lots of quantum and non-perturbative effects,
which potentially can not only affect the details, but also lead to new physics.
As one might anticipate, such effects are under better control in supersymmetric theories and in situations
where surface operators preserve some fraction of supersymmetry.


\subsection{Construction of surface operators}
\label{sec:definition}

Now, once we have presented the basic idea of what a surface operator is, we can elaborate on various points,
starting with the definition.
In the standard formulation of quantum field theory, based on a Feynman path integral,
there are several (often, equivalent) ways to define surface operators \cite{Gukov:2006jk,Gukov:2008sn}:

\begin{itemize}

\item
as singularities or boundary conditions for the gauge filed $A_{\mu}$ (and, possibly, other fields)
along a surface $D$ in four-dimensional space-time;

\item
as a coupled 2d-4d systems, namely a 2d theory supported on $D$ with a flavor symmetry group $G$
that is gauged upon coupling to 4d theory on $M$.

\end{itemize}
The latter option, in turn, is often subdivided into two large classes of models where the 2d theory on $D$ is either
$a)$ gauge theory itself, or $b)$ non-linear sigma-model. Clearly, these two classes do not exhaust all possibilities
and, yet, there are models which belong to both. A prominent example of such a model that has the advantage of being looked at from
several viewpoints is a 2d sigma-model with target space $\cp^1 = \C^2 / \!\! / U(1)$ that can be equivalently described
as a GLSM with $U(1)$ gauge group. It defines a surface operator in 4d gauge theory with gauge group $G = SU(2)$
that is a symmetry of $\cp^1$ (and for which $\C^2$ is the defining two-dimensional representation).

As for the first way of defining surface operators, we already saw examples in \eqref{etafirst} and \eqref{alphafirst}
where one did not need to introduce any additional 2d degrees of freedom. In particular, the disorder operator \eqref{alphafirst}
has an obvious analogue in a non-abelian gauge theory with a general gauge group $G$.
Namely, one can define operators supported on a surface $D$
by requiring the gauge field $A$ (and, possibly, other fields)
to have a prescribed singularity along $D$:
\begin{equation}
{\rm Hol}_{\ell} (A) \in \frak C
\label{holonomy}
\end{equation}
where $\ell$ is a small loop that links surface $D \subset M$ in the space-time 4-manifold $M$,
and $\frak C$ is a fixed conjugacy class in the gauge group $G$ (or, possibly, its complexification $G_{\C}$).
The latter option, $\frak C \subset G_{\C}$, is realized in $\CN \ge 2$ supersymmetric gauge theory,
where the gauge field $A$ combines with a Higgs field $\phi$ in a complex combination $\CA = A + i \phi$
(see {\it e.g.} Figure \ref{Hassefig} for a list of complex conjugacy classes in $SO(7)$ and $Sp(6)$ gauge theory).

\subsection{Classification of surface operators}
\label{sec:classification}

A careful definition of surface operators essentially
gives an answer to the question about classification of surface operators.

\begin{figure}[t] \centering \includegraphics[width=4.0in]{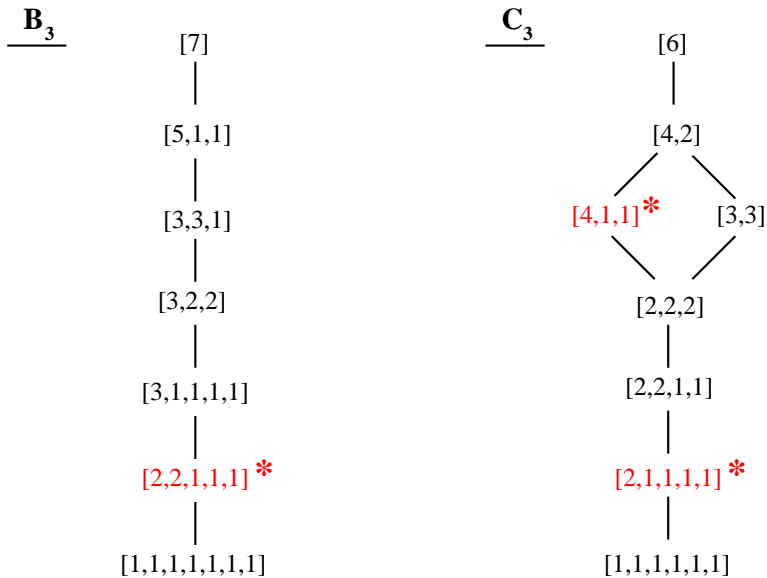}
\caption{Surface operators shown in red and labeled by $*$ appear to spoil electric-magnetic duality
between $SO(7)$ and $Sp(6)$ gauge theories.
In order to restore a nice match, one has to introduce a larger class of surface operators.
\label{Hassefig}} \end{figure}

In general, parameters of surface operators can be divided into
discrete data and continuous parameters.
In a way, the former is analogous to the choice\footnote{For example,
in $\CN=2^*$ theory with gauge group $G=SU(N)$ this choice
includes the choice of a partition of $N$. When $G$ is a classical group of Cartan type $B$, $C$, or $D$,
the choice of partition must satisfy certain conditions, as illustrated in Figure \ref{Hassefig}.
In particular, the transformation of surface operators under electric-magnetic duality becomes a rather non-trivial matter
in non-abelian gauge theories.} of a representation
that labels line operators, {\it cf.} (\ref{wilson}),
while the latter are a novel feature of surface operators.
Moreover, it turns out that understanding these continuous
parameters is the key to addressing other important questions
about the properties of surface operators.
For example, the non-commutative structure of line operators
supported on a surface operator --- that will be discussed in section \ref{sec:lines} ---
is described by the fundamental group \eqref{fundparam} of the suitable (sub)space of continuous parameters.

In our previous discussion we already saw examples of both discrete and continuous parameters.
In \eqref{aetaabelian}, the parameters $\alpha$ and $\eta$ are examples of continuous parameters,
whereas the choice of the subgroup $\LL \subseteq G$, called the Levi subgroup,
preserved by the surface operator along $D$ is a typical example of the discrete parameter.
Although we introduced these parameters in the simplest (abelian) examples,
they have immediate analogues in a very broad class of surface operators known at present.
The number of continues parameter can vary, usually from 0 to the rank of the gauge group $G$ (multiplied by $\CN$).
The surface operators which do not have continuous parameters at all are usually called {\it rigid}.

The classification problem consists of making a list of discrete and continues parameters that label surface operators in a given gauge theory.
At present, this is an open problem, which is very far from satisfactory solution.
One might hope to make more progress by imposing additional conditions,
{\it e.g.} focusing on SUSY gauge theories and requiring surface operators to preserve some fraction of supersymmetry.
Thus, one might hope that the general construction of \cite{Gukov:2008sn} is not too far from a complete classification
of half-BPS surface operators in the maximally supersymmetric gauge theory in four dimensions.
The next natural step is the classification of half-BPS surface operators in $\CN=2$ gauge theories (that will be of our main interest here),
followed by $\tfrac{1}{4}$-BPS surface operators and surface operators in $\CN=1$ and $\CN=0$ gauge theories.\footnote{See
\cite{Koh:2008kt,Gadde:2013wq} for discussion of $\tfrac{1}{4}$-BPS surface operators in 4d $\CN=2$ gauge theories.}

Let us illustrate how this construction and classification of half-BPS surface operators works in the simplest case,
namely in $\CN=4$ super-Yang-Mills, which can be viewed as a special case of $\CN=2$ gauge theory with
a massless hypermultiplet in the adjoint representation of the gauge group $G$.
(Its deformation by turning on the mass parameter $m \ne 0$ for the adjoint hypermultiplet is usually called $\CN=2^*$ theory.)
Much like in our preliminary discussion around \eqref{alphafirst},
we can produce a large class of half-BPS surface operators which break the gauge
group down to a Levi subgroup ${\Bbb{L}} \subset G$ and which also break the global $R$-symmetry group,
\begin{equation}
SO(6)_{R} \to SO(4) \times SO(2)
\label{rsymbreaking}
\end{equation}
by introducing a singularity for the gauge field that corresponds to the monodromy (\ref{holonomy})
and for two components of the Higgs field, say $\varphi = \phi_1 + i \phi_2$,
\bea
A & = & \alpha d \theta + \ldots \,, \label{asing} \\
\varphi & = & {1 \over 2} \big( \beta + i \gamma \big) {dz \over z} + \ldots \label{phising}
\eea
Here, $z = x^2 + i x^3 = r e^{i \theta}$ is a local complex coordinate,
normal to the surface $D \subset M$, and the dots stand for less singular terms.
In order to obey the supersymmetry equations \cite{Gukov:2006jk},
the parameters $\alpha$, $\beta$, and $\gamma$ must take values in
the ${\Bbb{L}}$-invariant part of $\frak t$, the Lie algebra of the maximal torus $\TT$ of $G$.
Moreover, gauge transformations shift values of $\alpha$ by elements of the cocharacter lattice, $\Lambda_{{\rm cochar}}$.
Hence, $\alpha$ takes values in $\TT = {\frak t} / \Lambda_{{\rm cochar}}$.

In addition to the classical (or ``geometric'') parameters
$(\alpha,\beta,\gamma)$, the surface operators of this type
are also labeled by quantum parameters \eqref{etafirst},
the ``theta angles'' of the two-dimensional theory on $D \subset M$.
It is easy to see that parameters $\eta$
take values in the $\Bbb{L}^{\vee}$-invariant part of
the maximal torus of the Langlands/GNO dual group $G^{\vee}$.
We can summarize all this by saying that maximally supersymmetric
($\CN=4$) super-Yang-Mills theory admits a large class of surface operators
labeled by a choice\footnote{In a theory with gauge group $G=SU(N)$
this choice is equivalent to a choice of a partition of $N$.}
of the Levi subgroup ${\Bbb{L}} \subset G$ and continuous
parameters
\begin{equation}
(\alpha,\beta,\gamma,\eta)
\; \in \; \big( \TT \times \frak t \times \frak t \times \TT^{\vee} \big) / \Weyl
\label{surfparameters}
\end{equation}
invariant under the Weyl group $\Weyl_{{\Bbb{L}}}$ of ${\Bbb{L}}$.
Similar surface operators exist in
$\CN=2$ super\-symmetric gauge theories; the only difference
is that they don't have parameters $\beta$ and~$\gamma$.

These surface operators naturally correspond to the so-called Richardson
conjugacy classes in the complexified gauge group $G_{\mathbb{C}}$,
{\it cf.} (\ref{holonomy}) and,
in a theory with gauge group $G=SU(N)$, cover all half-BPS
surface operators which correspond to singularities with simple poles.

\subsection{Surface operators in 4d $\CN=2$ gauge theory}
\label{sec:N2}

The construction described in the end of the previous section can be easily generalized to
define half-BPS surface operators in $\CN=2$ gauge theories,
see {\it e.g.} \cite{Gukov:2007ck,Koh:2009cj,Tan:2009iu,Alday:2009fs,Gaiotto:2009fs}
and subsequent work.
As a result, one finds a fairly large class of surface operators labeled by the Levi subgroup $\LL \subseteq G$
and continuous parameters $(\alpha,\eta) \in \big( \TT \times \TT^{\vee} \big) / \Weyl$, which in $\CN=2$ theories
conveniently unify into holomorphic combinations
\be
t = \eta + \tau \alpha
\ee
where $\tau$ is the coupling (matrix) of the $\CN=2$ gauge theory.

A novel feature of such half-BPS surface operators in $\CN=2$ theories
--- compared to maximally supersymmetric Yang-Mills or abelian ($\CN=0$) theories without supersymmetry discussed above ---
is that one must be wary of quantum effects,
which can not only renormalize the values of various parameters but also change the nature of a surface operator altogether.
In other words, defined as a singularity for the gauge field (and, possibly, other fields) as described in section \ref{sec:definition},
a surface operator is defined at a given energy scale in the 4d theory.
It can be a UV theory, or an IR theory, or some effective theory at intermediate energy scale.
An interesting question, then, is to study what becomes of such surface operator at other energy scales and / or regimes of parameters.

In order to answer such questions, it is often helpful to use another definition of surface operators described in section \ref{sec:definition}.
Namely, one can define a surface operator supported on $D \subset M$ by introducing additional 2d degrees of freedom along $D$,
with their own Lagrangian and a global symmetry group $G$ that becomes gauged upon coupling to 4d degrees of freedom.
Of course, if 4d gauge theory in question has matter fields $Q$, they too can be coupled to 2d degrees of freedom supported on $D$
in a gauge invariant manner. As explained in \cite{Gukov:2006jk,Gukov:2008sn}, integrating out 2d degrees of freedom
leaves behind a singularity (obviously, supported on $D$) in the field equations of the four-dimensional theory:
\bea
F_{23} - Q Q^{\dagger} & = & 2\pi \delta^2 (\vec x) \mu_1  \label{FQsing} \\
D_{\bar z} Q & = & \pi \delta^2 (\vec x) (\mu_2 + i \mu_3) \nonumber
\eea
where $\CN=4$ SYM corresponds to a special case when $Q$ transforms in the adjoint representation of the gauge group $G$.

Which two-dimensional theories can one use in this construction?
In general, any 2d theory will do as long as it has a symmetry group $G$ that can be gauged and as long as it is free of anomalies.
In fact, coupling 2d degrees of freedom to 4d gauge theory even allows one to experiment with anomalous 2d theories
where anomalies can be canceled by the inflow from the four-dimensional bulk \cite{Callan:1984sa}.

When one aims to build a surface operator that preserves certain symmetries of the four-dimensional gauge theory,
the 2d theory on the defect must be chosen accordingly, so that it also enjoys the desired symmetries.
For instance, if the goal is to build a half-BPS surface operator in a supersymmetric 4d gauge theory,
the 2d theory on $D$ must have at least half of the supercharges present in 4d, as illustrated in Table~\ref{tab:2d4dtarget}.

A simple way to achieve this is to take 2d theory to be a sigma-model with the desired supersymmetry and a target space $X$
that has a symmetry group $G$. Of course, depending the on the desired amount of supersymmetry, the space $X$ may also need to be
K\"ahler or hyper-K\"ahler for applications to $\CN=2$ and $\CN=4$ gauge theory, respectively.
Large class of such targets that have all the desired properties are coadjoint orbits
(or, via the exponentiation map, conjugacy classes $X = \frak C$) and their complexifications.
Indeed, they admit K\"ahler and hyper-K\"ahler metrics, respectively, in addition to a $G$-action
that one needs for coupling to 4d degrees of freedom.
In the $\CN=2$ case, let $\mu_1$ be the moment map for the action of $G$ on the K\"ahler target space $X$ and,
similarly, in the $\CN=4$ theory let $\vec \mu = (\mu_1, \mu_2, \mu_3)$ be the hyper-K\"ahler moment map for the action of $G$ on $X$.

Then, integrating out 2d degrees of freedom in these cases leads to surface operators defined as singularities \eqref{FQsing},
where the holonomy of the (complexified) gauge field is required to be in a fixed conjugacy class, {\it cf.} \eqref{holonomy}.
This provides a link between two ways of defining surface operators described in section \ref{sec:definition},
namely, as singularities and as coupled 2d-4d systems.

Supersymmetry also often tightly constrains the geometry of the surface $D \subset M$.
A popular example is $D = \R^2$ linearly embedded in $M = \R^4$ which breaks the Lorentz symmetry as, {\it cf.} \eqref{224}:
\be
SO(1,3) \; \to \; SO(1,1)_{01} \times SO(2)_{23}
\label{SOSOSO}
\ee
where, for concreteness, we chose the surface operator to be oriented along the $(x^0,x^1)$ plane.
Since surface operators break Lorentz symmetry, they must break at least part of the supersymmetry
and some of the $R$-symmetries. Thus, in $\CN=4$ gauge theory the $R$-symmetry breaking pattern is \eqref{rsymbreaking}.
Similarly, a generic $\CN=2$ gauge theory has $R$-symmetry group $SU(2)_R \times U(1)_r$,
of which $U(1)_r$ may be broken by quantum effects. A half-BPS surface operator further breaks $SU(2)_R$ down to $U(1)_R$.

Of particular interest to us, especially in applications to the AGT correspondence \cite{Alday:2009aq}
will be half-BPS surface operators in superconformal gauge theories.
The conformal group in four dimensions is $SO(4,2) \sim SU(2,2)$, and a surface operator oriented along
the $(x^0,x^1)$ plane breaks it down to a subgroup, {\it cf.} \eqref{SOSOSO}:
\be
SO(2,2) \times U(1)_{23} \subset SO(4,2)
\label{confsurface}
\ee
Here, $SO(2,2)\cong SL(2,\R)_{\ttL}\times SL(2,\R)_{\ttR}$ is the conformal group in two dimensions and $U(1)_{23}$
is the rotation symmetry in the $(x^{2},x^{3})$ plane transverse to the surface operator.

The analogous symmetry breaking patterns in supersymmetric theories are summarized in Table~\ref{tab:2d4dtarget}.
In particular, the superconformal symmetry group of 4d $\CN=2$ gauge theory is $SU(2,2|2)$.
Its bosonic subgroup is $S[U(2,2)\times U(2)]\sim SU(2,2)\times SU(2)_R\times U(1)_r$,
where $SU(2,2)$ is the familiar conformal group and $SU(2)_R \times U(1)_r$ is the $R$-symmetry group mentioned earlier.
Apart from the conformal symmetry \eqref{confsurface}, a half-BPS surface operator also preserves
$U(1)_{\ttL} \times U(1)_{\ttR} \subset SU(2)_{R}\times U(1)_{r}$ part of the $R$-symmetry group
and four (out of eight) supercharges $\CQ^2_-,{\tilde \CQ}^1_{\dot -},\CQ^1_+,{\tilde \CQ}^2_{\dot +}$
of the four dimensional theory.
The bosonic subgroup $SL(2,\R)_{\ttL}\times U(1)_\ttL$ combines with the supersymmetries $\CQ^2_-,{\tilde \CQ}^1_{\dot -}$
to form $SU(1,1|1)_\ttL$. Similarly, the remaining charges generate $SU(1,1|1)_\ttR$,
so that in total a half-BPS surface operator in $\CN=2$ superconformal theory preserves
$SU(1,1|1)_\ttL\times SU(1,1|1)_\ttR \times U(1)_e$ subgroup of $SU(2,2|2)$, where $U(1)_e$ is the commutant of the embedding.

\begin{table}
\begin{centering}
\begin{tabular}{|c|c|c|}
\hline
~4d theory on $M$~ & ~2d theory on $D$~ & ~superconformal symmetry~ \tabularnewline
\hline
\hline
$\CN=4$ & $\CN=(4,4)$ & $PSU(2,2|4) \to PSU(1,1|2) \times PSU(1,1|2) \times U(1)$ \tabularnewline
\hline
$\CN=2$ & $\CN=(2,2)$ & $SU(2,2|2) \to SU(1,1|1) \times SU(1,1|1) \times U(1)$ \tabularnewline
\hline
$\CN=1$ & $\CN=(0,2)$ & $SU(2,2|1) \to SU(1,1|1) \times SL(2,\R) \times U(1)$ \tabularnewline
\hline
\end{tabular}
\par\end{centering}
\caption{\label{tab:2d4dtarget}Half-BPS surface operators in SUSY gauge theories can be described
as coupled 2d-4d systems with suitable amount of supersymmetry in 2d theory.
The last column is only relevant to superconformal theories and describes the symmetry breaking pattern due to surface operator.}
\end{table}

\subsection{Their role in AGT correspondence}
\label{sec:AGT}

Now we are ready to review the role of surface operators in the 2d-4d correspondence \cite{Alday:2009aq}
that relates Liouville conformal block on a Riemann surface $C$ and the equivariant instanton partition function \cite{Nekrasov:2002qd}
of the class $\CS$ gauge theory \cite{Gaiotto:2009we,Gaiotto:2009hg} labeled by the Riemann surface $C$:
\bea
Z^{\text{inst}} (a,\tau,\epsilon) & = & Z^{\text{Liouv}} (\alpha,q,b) \nonumber \\
\epsilon_1 : \epsilon_2  & = & b : 1/b \label{AGTdict} \\
\exp (2\pi i \tau_{\text{UV}}) & = & q \nonumber \\
a & = & \alpha - Q/2 \nonumber
\eea
where $Q = b+ 1/b$ is the standard notation in the literature on Liouville theory.
Note, the left-hand side of this dictionary involves supersymmetric gauge theory in four dimensions, whereas the right-hand side
is about a {\it non-supersymmetric} 2d theory.
There is a similar version of this correspondence \cite{Gadde:2011uv}
that relates superconformal index of the 4d $\CN=2$ gauge theory $\CT [C]$ labeled by $C$
and a certain deformation of 2d non-supersymmetric Yang-Mills theory on $C$ which we shall briefly discuss in section \ref{sec:index}.

The conformal block in \eqref{AGTdict} represents a ``chiral half'' of the full Liouville correlation function,
which has the form of an integral of the absolute value squared of a conformal block
and also admits a nice interpretation in 4d gauge theory as a partition function \cite{Pestun:2007rz} on a 4-sphere $S^4$.
Indeed, dividing $S^4$ into the northern and southern hemispheres illustrated in Figure~\ref{stereo}
corresponds to the chiral decomposition of the Liouville CFT correlation functions into ``left-moving'' and ``right-moving'' chiral halves.

\FIGURE{
\includegraphics[width=2in]{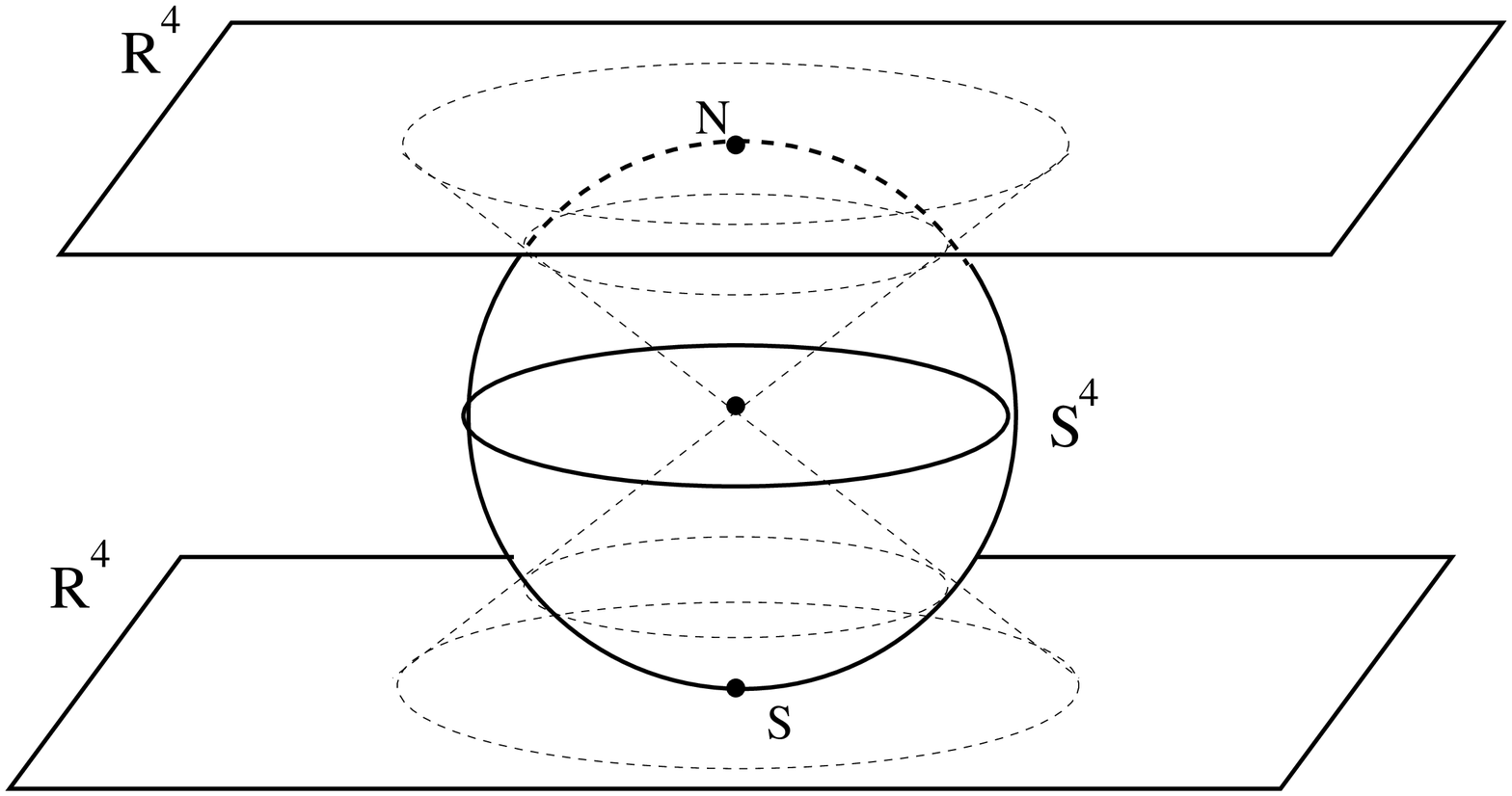}
\caption{Upon the hemispherical stereographic projection on two copies of $\R^4$, surface operators on $S^4$ factorize into a two surface operators, a north and a south half, glued together at the equator.\\}
\label{stereo}
}

The $\Omega$-deformation of the Euclidean $\CN=2$ gauge theory on $M = \R^4$,
used in the definition of the instanton partition function, involves the subgroup of the rotation symmetry
\be
SO(2) \times SO(2) \subset SO(4)
\ee
which is precisely the part of the symmetry preserved by a surface operators, {\it cf.} \eqref{SOSOSO}.
Therefore, following \cite{Nekrasov:2002qd}, one can introduce $\Omega$-deformation and the partition function
in the presence of a surface operators:
\be
\label{zkmdef}
Z^{{\rm inst}}_{k,{\mathfrak m}} (\epsilon_1, \epsilon_2) = \oint_{\CM_{k,{\mathfrak m}}} 1
\ee
Here, $\CM_{k,{\mathfrak m}}$ is the moduli space of ``ramified instantons'' on $M \setminus D$
labeled by the ordinary instanton number $k := c_2 (E)$ 
and the monopole number
\begin{equation}
\label{mnumber}
{\mathfrak m} = \frac{1}{2\pi} \int_{D} F \qquad\qquad {\rm (``monopole~number'')}
\end{equation}
that measures the magnetic charge of the gauge bundle $E$ restricted to $D$.
Then, the path integral of the $\Omega$-deformed $\CN=2$ gauge theory
in the presence of a surface operator of Levi type $\LL$ gives the generating function
\begin{equation}
\label{zinstkm}
Z^{{\rm inst}} (a,\Lambda,\epsilon;\LL,z)
= \sum_{k=0}^{\infty} ~\sum_{{\mathfrak m} \in \Lambda_{\LL}}~
\Lambda^{2Nk} e^{i z \cdot {\mathfrak m}} ~Z^{{\rm inst}}_{k,{\mathfrak m}} (a,\epsilon) 
\end{equation}
where the coefficients $Z^{{\rm inst}}_{k,{\mathfrak m}}$ are precisely the integrals \eqref{zkmdef}.

The basic surface operator (with next-to-maximal $\LL = S[U(1) \times U(N-1)]$) is labeled by a single complex parameter $z = \eta + i \alpha$
that takes values in $C$. Incorporating this surface operator in the instanton partition function on $\R^4_{\epsilon_1,\epsilon_2}$
or $S^4_{\epsilon_1,\epsilon_2}$ on the Liouville side corresponds to inserting a degenerate primary operator at a point $z \in C$,
\be
\Phi_{2,1} (z) = e^{-(b/2) \phi (z)}
\ee

There are several ways to argue for this identification:

\begin{itemize}

\item
using higher-dimensional constructions (that will be discussed below),

\item
using the ``semi-classical limit'' $\epsilon_{1,2} \to 0$,

\item
studying line operators within the surface operator,

\item
using tests based on direct computations of both sides.

\end{itemize}

\noindent
Since higher-dimensional constructions and line operators will be discussed in sections \ref{sec:higher} and \ref{sec:lines},
respectively, let us make a few comments on the semi-classical limit $\epsilon_{1,2} \to 0$.
One of the main results of \cite{Nekrasov:2002qd} is that, in this limit, the (logarithm of the) instanton partition
function has a second order pole, whose coefficient is the Seiberg-Witten prepotential $\CF (a_i)$.
This matches the structure of the Liouville conformal block in the limit $\hbar^2 = \epsilon_1 \epsilon_2 \to 0$.
The insertion of a degenerate field does not affect the leading singularity, but leads to a new first-order pole
\be
Z^{\text{Liouv}} \sim \exp \left( - \frac{\CF (a_i)}{\hbar^2} + \frac{b \CW (a_i,z)}{\hbar} + \ldots \right)
\ee
which has an elegant translation to the language of 4d $\CN=2$ gauge theory.
\FIGURE[l]{
\includegraphics[width=2in]{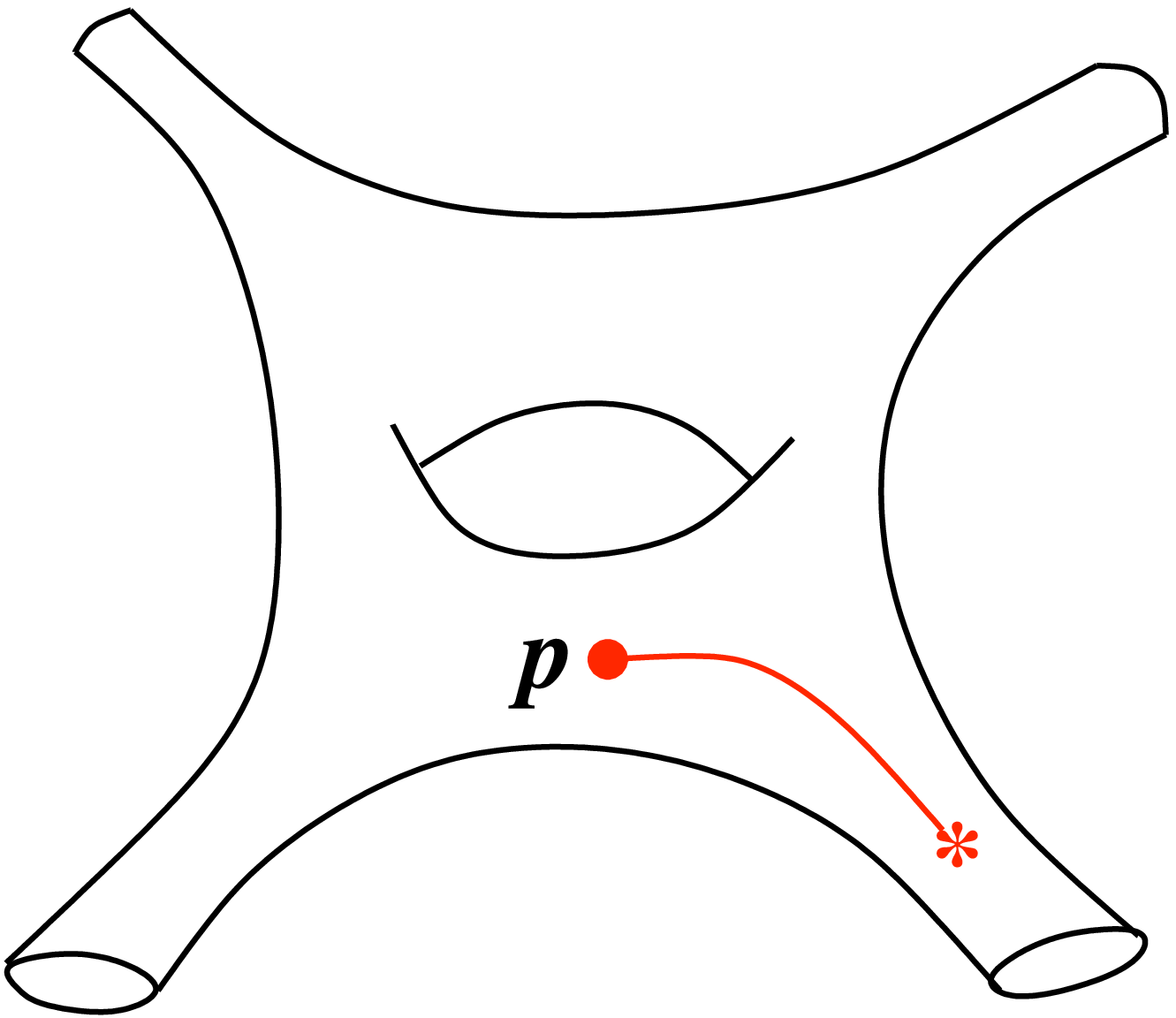}
\caption{The effective twisted superpotential $\CW$ can be expressed as an integral over an open path on the Seiberg-Witten curve $\Sigma$.\\}
\label{intfig}
}
Indeed, using basic properties of the AGT correspondence,
one can identify the function $\CW (a_i,z)$ with the an integral \cite{Alday:2009fs}:
\be
\CW = \int_{p_*}^p \lambda_{SW}
\label{swint}
\ee
along some path on the Seiberg-Witten curve, starting at some reference point $p_*$ (see Figure \ref{intfig}).
This is precisely how the insertion of a surface operator modifies the instanton partition function:
\be
Z^{\text{inst}} \sim \exp \left( - \frac{\CF (a_i)}{\epsilon_1 \epsilon_2} + \frac{\CW (a_i,t)}{\epsilon_1} + \ldots \right)
\ee
The leading singularities here come from the ``regularized'' equivariant volume contributions of the 4d bulk degrees
of freedom supported on $M = \R^4_{\epsilon_1,\epsilon_2}$ and the 2d contribution of a surface operator supported on $D = \R^2_{\epsilon_1}$:
\be
\text{Vol} (\R^4_{\epsilon_1,\epsilon_2}) = \int_{\R^4_{\epsilon_1,\epsilon_2}} 1 = \frac{1}{\epsilon_1 \epsilon_2}
\qquad , \qquad
\text{Vol} (\R^2_{\epsilon_1}) = \int_{\R^2_{\epsilon_1}} 1 = \frac{1}{\epsilon_1}
\ee
Naturally, here we are more interested in the contribution of a surface operator.
The function $\CW (a_i,z)$ that depends on both the Coulomb branch parameters of the 4d theory as well as
continuous parameters of the surface operator has a simple physical interpretation:
it is the effective twisted superpotential of the 2d $\CN=(2,2)$ theory on $D$.
The relation
\be
d \CW = \eta da + \alpha d a_D
\ee
tells us that the IR parameters $(\alpha,\eta)$ coincide with the points on the Jacobian of the Seiberg-Witten curve.
Indeed, differentiating with respect to $a$,
\be
\partial_a \CW = \eta + \tau \alpha
\ee
and using \eqref{swint} we conclude that the map
\be
t_i = \frac{\partial \CW}{\partial a^i} = \int_{p_*}^p \frac{\partial \lambda}{\partial a^i}
= \int_{p_*}^p \omega_i
\ee
is precisely the Abel-Jacobi map from a Riemann surface to its Jacobian.

Note, the shifts of $\CW$ by $n_e a + n_m a_D$ correspond to the monodromies of $\alpha$ and $\eta$.
In section \ref{sec:lines} we relate them to line operators localized within a surface operator.




\section{Surface operators from higher dimensions}
\label{sec:higher}

Four-dimensional $\CN=2$ gauge theories --- that are of our prime interest here in view of the AGT correspondence ---
can be realized in a variety of higher-dimensional models, that include 6d $(0,2)$ theory, type II string theory, and M-theory.
Even though such constructions involve more sophisticated higher-dimensional systems, they often shed light on
strongly coupled gauge dynamics and help understand various aspects of $\CN=2$ gauge theories, such as the Seiberg-Witten exact
solution \cite{Seiberg:1994rs,Seiberg:1994aj} and Nekrasov's (K-theoretic) instanton partition function~\cite{Nekrasov:2002qd}.

For example, a nice heuristic derivation of the AGT correspondence \eqref{AGTdict} follows from the $(0,2)$ superconformal
theory in six dimensions, which combines the 2-manifold $C$ (where the Liouville theory lives) and the 4-manifold
$M=\R^4_{\epsilon_1,\epsilon_2}$ (where the 4d $\CN=2$ gauge theory lives):
$$
\begin{array}{ccccc}
\; & \; & \text{6d $(0,2)$ theory} & \; & \; \\
\; & \; & \text{ on $C \times \R^4_{\epsilon_1,\epsilon_2}$} & \; & \; \\
\; & \swarrow & \; & \searrow & \; \\
\text{2d Liouville theory} & \; & \; & \; & \text{4d $\CN=2$ theory $\CT[C]$} \\
\text{of $C$} & \; & \; & \; & \text{on $\R^4_{\epsilon_1,\epsilon_2}$}
\end{array}
$$
Here, the two sides of the AGT correspondence \eqref{AGTdict} are simply the two ways of reducing the 6d theory,
either on a 2-manifold $C$ or on a 4-manifold $M=\R^4_{\epsilon_1,\epsilon_2}$ (or $M=S^4_{\epsilon_1,\epsilon_2}$).
In order to preserve supersymmetry, the former must be accompanied by a partial topological twist~\cite{Bershadsky:1995qy},
whereas the latter involves deformed supersymmetry algebra that can be conveniently understood via coupling
to the corresponding off-shell supergravity theory~\cite{Festuccia:2011ws}.

The 6d $(0,2)$ theory itself admits surface operators (a.k.a. codimension-4 defects)
which, upon reduction on $C$, give rise to surface operators in 4d $\CN=2$ theory $\CT [C]$.
The existence of such surface operators can be deduced by realizing 6d $(0,2)$ theory itself on the world-volume of
$N$ five-branes supported on $C \times M \times \{ \text{pt} \}$ in 11d M-theory on $T^* C \times M \times \R^3$.
And, in order to reduce to a surface operator on $M$, the codimension-4 defect of the six-dimensional
theory must be supported at a point on $C$. From the viewpoint of the 4d gauge theory, its position $z \in C$
becomes a continuous parameter that labels half-BPS surface operator.

Note, six-dimensional $(0,2)$ superconformal theory also has codimension-2 defects that can also produce
half-BPS surface operators in four dimensions upon wrapping all of the Riemann surface $C$ \cite{Alday:2010vg}.
Since codimension-2 defects carry $G$-bundles, such surface operators are naturally labeled
by points $x \in \text{Bun}_G (C)$.
These surface operators are dual to the surface operators that arise from codimension-4 defects \cite{toappear}.

\subsection{Brane constructions}
\label{sec:branes}

In addition, there exist various string constructions of surface operators.
The one relevant to our discussion here is based on the brane realization of
$\CN=2$ gauge theory in type IIA string theory \cite{Witten:1997sc},
where basic surface operators (with next-to-maximal $\LL$)
can be described by introducing semi-infinite D2-branes \cite{Alday:2009fs}:
\begin{align}\label{iiabranes}
\hbox{NS5} &: \quad 012345 \cr
\hbox{D4}  &: \quad 0123~~~6 \cr
\hbox{D2}  &: \quad 01~~~~~~~~7
\end{align}
Lifting this configuration to M-theory, we obtain a M5-brane
with world-volume $\R^4 \times \Sigma$ and a M2-brane (ending on the M5-brane)
with world-volume $\R^2 \times \R_+$.
Here, $D = \R^2$ is the support of the surface operator in
the four-dimensional space-time $M = \R^4$, and $\Sigma$ is the Seiberg-Witten curve
of the $\CN=2$ gauge theory.

\begin{figure}[t] \centering \includegraphics[width=5.0in]{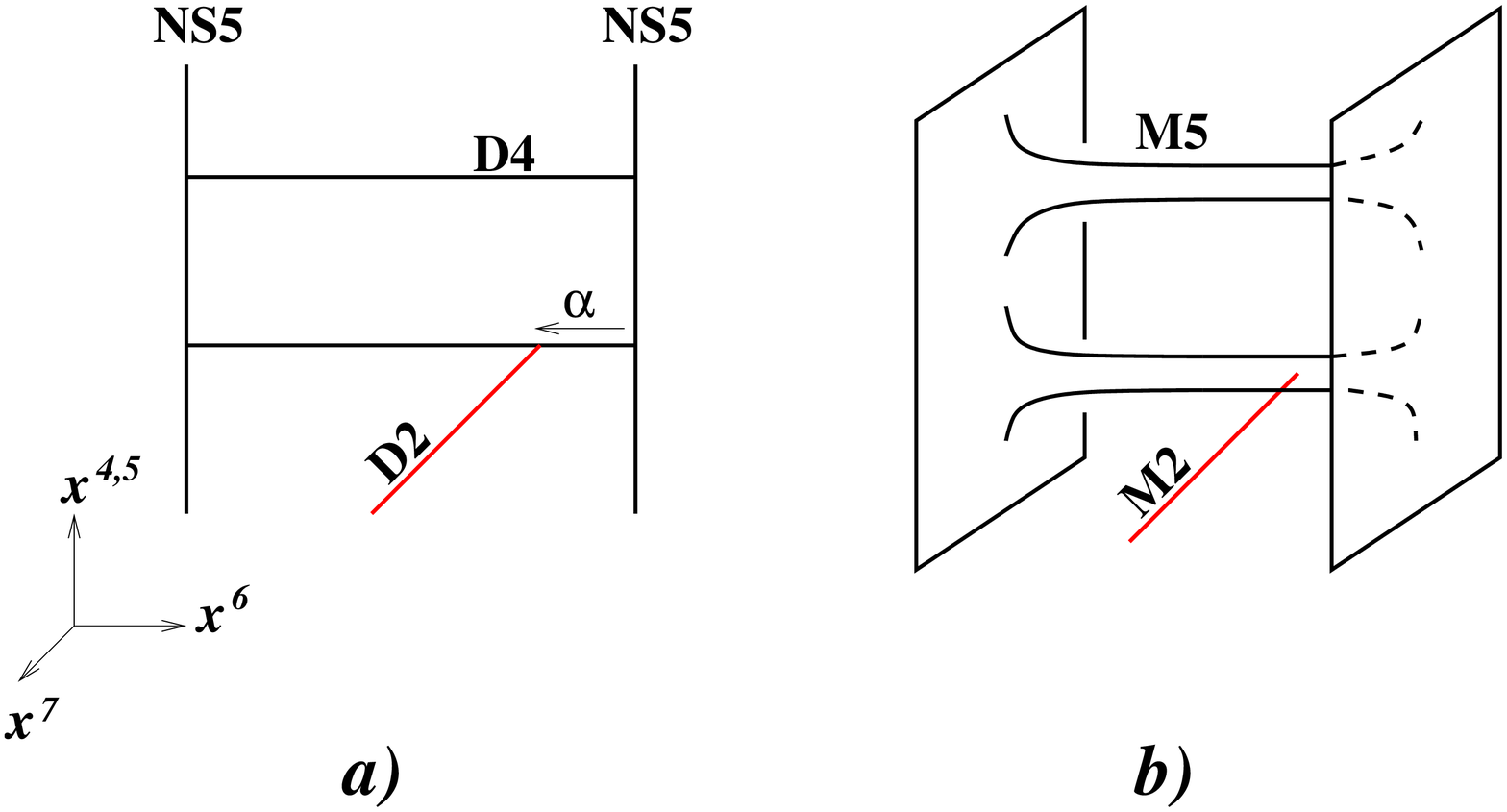}
\caption{The brane construction of $\CN=2$ super Yang-Mills theory
with a half-BPS surface operator in type IIA string theory $(a)$ and its M-theory lift $(b)$.
\label{branefig}} \end{figure}

In this construction, the M2-brane is localized along $\Sigma$
(the choice of the point $t \in \Sigma$ corresponds to the IR parameters
of the surface operator) and has a semi-infinite extent along the direction $x^7$,
as described in \eqref{iiabranes}.

Similar construction can be used to define UV surface operators in
4d $\CN=2$ superconformal theories obtained from compactifications of 6d $(0,2)$ fivebrane theory on a UV Riemann surface $C$.

\subsection{Geometric engineering}
\label{sec:geomeng}

Let us consider a four-dimensional $\CN=2$ gauge theory that can be geometrically
engineered via type IIA string ``compactification'' on a Calabi-Yau space $CY_3$.
In other words, we take the ten-dimensional space-time to be $M \times CY_3$,
where $M$ is a 4-manifold (where $\CN=2$ gauge theory lives) and $CY_3$ is a suitable Calabi-Yau space \cite{Katz:1996fh}.
We recall that such $CY_3$ is non-compact
and toric, and that its toric polygon coincides with the Newton polygon of the Seiberg-Witten curve $\Sigma$.
As in most of our applications, one can simply take $M = \R^4$.

Aiming to reproduce half-BPS surface operators supported on $D = \R^2$,
we need an extra object that breaks part of the Lorentz symmetry (along $M = \R^4$)
and half of the supersymmetry.
It is easy to see that D4-branes supported on supersymmetric 3-cycles in $CY_3$
provide just the right candidates \cite{Ooguri:1999bv}.
Indeed, if the world-volume of a D4-brane is $\R^2 \times L$, where
\be
\begin{matrix}
{\mbox{\rm space-time:}} & \qquad & \R^4 & \times & CY_3 \\
& \qquad & \cup &  & \cup \\
{\mbox{\rm D4-brane:}} & \qquad & \R^2 & \times & L
\end{matrix}
\label{surfeng}
\ee
and $L$ is a special Lagrangian submanifold of $X$,
then such a D4-brane preserves exactly the right set of symmetries
and supersymmetries as the half-BPS surface operators discussed in section~\ref{sec:N2}.

\FIGURE[l]{
\includegraphics[width=2in]{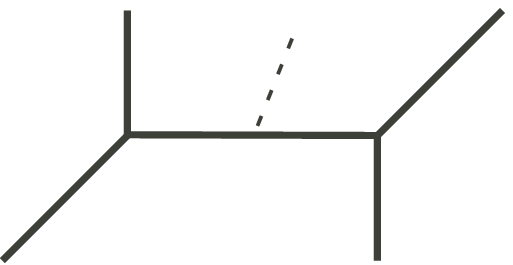}
\caption{ $U(1)$ toric geometry with a single Lagrangian brane.\\}
\label{fig:U1basic}
}

A nice feature of this construction is that it is entirely geometric:
all the parameters of a surface operators (discrete and continuous)
are encoded in the geometry of $L \subset CY_3$.
In particular, among the different choices of $L$ we should be
able to find those which correspond to half-BPS surface operators of Levi type $\LL$
with the continuous parameters $\alpha$ and $\eta$.
When $\LL = S[U(1) \times U(N-1)]$ is the next-to-maximal Levi subgroup,
the corresponding surface operator is geometrically engineered \cite{Dimofte:2010tz}
by a simple Lagrangian submanifold $L \cong S^1 \times \R^2$ invariant under the toric symmetry of $CY_3$
(see Figure \ref{fig:U1basic}).

The space of IR parameters of such a surface operator is the algebraic curve $\Sigma$ which is mirror
to the original Calabi-Yau 3-fold $CY_3$ via local mirror symmetry.\footnote{In the case of non-compact toric Calabi-Yau 3-folds
mirror symmetry (often called ``local mirror symmetry'') relates enumerative invariants of $CY_3$ with complex geometry of
a Riemann surface $\Sigma$.}
Equivalently, from the viewpoint of $CY_3$ these continuous parameters are open string moduli of $L$
corrected by world-sheet disk instantons.
Just like open string moduli become parameters of the surface operator in 4d $\CN=2$ gauge theory,
the gauge theory itself is determined by closed string moduli, which are K\"ahler parameters of $CY_3$.
(Note, a non-compact toric Calabi-Yau 3-fold $CY_3$ is rigid, {\it i.e.} has no complex structure deformations.)

\begin{equation}
\renewcommand{\arraystretch}{1.3}
\begin{tabular}{|@{\quad}c@{\quad}|@{\quad}c@{\quad}| }
\hline  {\bf 4d gauge theory} & {\bf geometry of} $CY_3$ \\
\hline
\hline
parameters of 4d $\CN=2$ theory & closed string moduli \\
($a_i$, $m_i$, $\Lambda$) & (K\"ahler moduli $Q$) \\
\hline
$\Omega$-background & string coupling / graviphoton \\
$\epsilon_1$ and $\epsilon_2$ & $q_1 = e^{\epsilon_1}$ and $q_2 = e^{\epsilon_2}$ \\
\hline
surface operator & Lagrangian submanifold \\
\hline
parameters of surface operator & open string moduli \\
\hline
$\Lambda_{\LL}$ & $H_1 (L; \Z)/$torsion \\
\hline
\end{tabular}
\label{closed_params}
\end{equation}

The geometric realization of a half-BPS surface operator \eqref{surfeng}
allows to express many interesting partition functions in terms enumerative invariants of the pair $(CY_3,L)$.
For example, the instanton partition function of the 4d $\CN=2$ gauge theory relevant to the AGT correspondence \eqref{AGTdict}
and its variant with a half-BPS surface operator \eqref{zinstkm}
both find a natural home on the right-hand side of the dictionary \eqref{closed_params} as so-called ``closed'' and ``open''
BPS partition functions, respectively.
To be more precise, the K-theoretic instanton counting on $M$ is captured by counting {\it refined BPS invariants} on $CY_3$:
\be
Z_{K}^{\rm inst}(\Lambda,a_i;q_1,q_2) = Z_{\rm BPS}^{\rm closed}(Q;q_1,q_2) \,.
\ee
Similarly, in the presence of a surface operator the K-theoretic analogue of \eqref{zinstkm} is equal to the generating
function of open (as well as closed) refined BPS invariants of the pair $(CY_3,L)$:
\be
Z^{\rm inst}_{\rm K-theory}(\Lambda,a_i,z;q_1,q_2) = Z^{\rm BPS}(Q,z;q_1,q_2) \,.
\label{BPSinst}
\ee
An important special case of this relation is the limit $\Lambda\to 0$ (\emph{i.e.}\ $Q_\Lambda\to 0$) and $(q_1,q_2)\to (q,1)$.
On the gauge theory side, this decouples the four-dimensional theory from the surface operator,
and counts vortices on the surface operator with respect to two-dimensional rotations (but not R-charge).
The resulting partition function counts only 2d vortices and not 4d instantons:
\be
Z^{\rm vortex}_{\rm K-theory}(z,a_i;q) = Z^{\rm open}_{\rm BPS}(Q_\Lambda=0,Q_{a_i},z;q,1) \,.
\label{BPSvortex}
\ee
For example, in the case of $CY_3$ shown in Figure~\ref{fig:U1basic} that engineers $U(1)$ gauge theory,
this limit corresponds to a degeneration upon which $(CY_3,L)$ is replaced by $(\C^3,L)$.


\subsection{Surface operators and BPS states}
\label{sec:BPS}

Upon lift to M-theory, the BPS states in the system \eqref{surfeng} are represented by membranes,
with and without boundary, as illustrated in \eqref{clcl} and \eqref{opclosed} below.
They are completely localized along $M$ and besides their support
in $CY_3$ have non-trivial extent only along the ``eleventh'' dimension of M-theory,
which can be treated as ``time'':
\be
\begin{matrix}
{\mbox{\rm space-time:}} & \qquad & \R & \times & M & \times & CY_3 \\
&  & \Vert & & \cup &  & \cup \\
{\mbox{\rm M5-brane:}} & \qquad & \R & \times & D & \times & L \\
&  & \Vert & & \cup &  &  \\
{\mbox{\rm M2-brane:}} & \qquad & \R & \times & \{ {\rm pt} \} & \times & \Sigma_g
\end{matrix}
\label{surfgeom}
\ee
In the five-dimensional gauge theory on $\R \times M$ such BPS states (open or closed) all look like particles.
Therefore, one can equivalently talk about BPS particles in 5d theory with a surface operator supported on $\R \times D$.
This system is very similar to our original 2d-4d system and can be related to that via reduction along one of the dimensions of $M$.
For either system, one can introduce the space of BPS states that can move in 4d (resp. 5d) bulk as well as the space of BPS states
localized on 2d (resp. 3d) surface operator, $\CH_{\text{BPS}}^{\text{bulk}}$ and $\CH_{\text{BPS}}^{\text{surface}}$.
The space $\CH_{\text{BPS}}^{\text{bulk}}$ depends only on the 4d/5d gauge theory on $M$ (resp. $\R \times M$),
whereas the space $\CH_{\text{BPS}}^{\text{surface}}$ depends on both 4d/5d gauge theory as well as the surface operator.

What is the relation between $\CH_{\text{BPS}}^{\text{bulk}}$ and $\CH_{\text{BPS}}^{\text{surface}}$?
The geometric engineering \eqref{surfgeom} can teach us an important lesson and help to answer this question.
It has been known for a long time that $\CH_{\text{BPS}}^{\text{bulk}}$ form an algebra \cite{Harvey:1996gc},
and recently it was further conjectured \cite{Gukov:2011ry} that the space of BPS states localized on a surface operator
forms a representation of this algebra
\be
\boxed{\phantom{\int}
\begin{array}{rcl}
\text{BPS states on a surface operator}: &  & \CH_{\text{BPS}}^{\text{surface}} \\
 & & \circlearrowleft \\
\text{BPS states in 4d/5d gauge theory}: &  & \CH_{\text{BPS}}^{\text{bulk}}
\end{array}
\phantom{\int}}
\label{HHrep}
\ee
Indeed, the space of BPS states in bulk gauge theory is graded by a charge lattice $\Gamma$,
which in the context of geometric engineering can be identified with even cohomology of local toric Calabi-Yau manifold:
\be
\Gamma \; = \; H^{\text{even}} (CY_3;\Z) \,.
\ee
Then, as explained in \cite{Harvey:1996gc},
two BPS states of the bulk theory, $\CB_1$ and $\CB_2$, of charge $\gamma_1, \gamma_2 \in \Gamma$
can form a bound state, $\CB_{12}$ of charge $\gamma_1 + \gamma_2$, as a sort of ``extension'' of $\CB_1$ and $\CB_2$,
\be
0 \; \to \; \CB_2 \; \to \; \CB_{12} \; \to \; \CB_1 \; \to \; 0 \,,
\label{extBBB}
\ee
thereby defining a product on $\CH_{\text{BPS}}^{\text{bulk}}$:
\be
\begin{array}{ccl}
\CH_{\text{BPS}}^{\text{bulk}} \; \otimes \; \CH_{\text{BPS}}^{\text{bulk}} & \longrightarrow & \CH_{\text{BPS}}^{\text{bulk}} \;  \\[.4cm]
(\; \CB_1 \;, \;\; \CB_2 \;) & \mapsto & \CB_{12}
\end{array}
\label{clcl}
\ee
$$
{\,\raisebox{-.5cm}{\includegraphics[width=5cm]{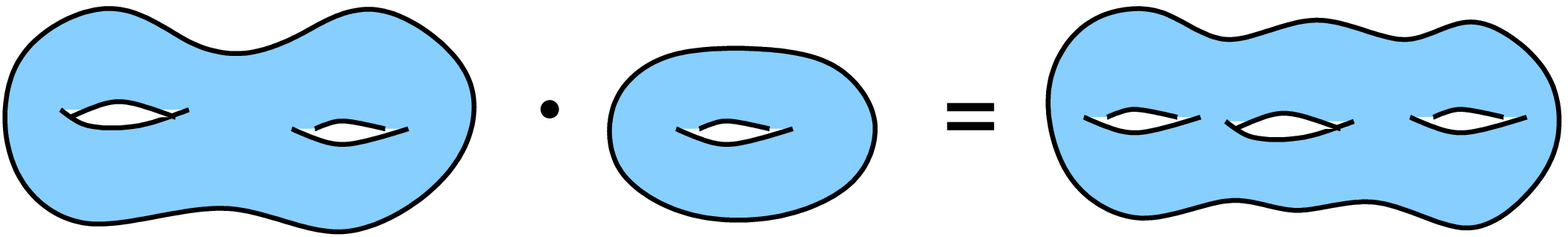}}\,}
$$
Mathematical candidates for the algebra $\CH_{\text{BPS}}^{\text{bulk}}$ include variants of the Hall algebra \cite{Schiffmann},
which by definition encodes the structure of the space of extensions \eqref{extBBB}:
\be
[\CB_1] \cdot [\CB_2] \; = \; \sum_{\CB_{12}} |0 \to \CB_2 \to \CB_{12} \to \CB_1 \to 0| \; [\CB_{12}]
\ee
In the present case, the relevant algebras include the motivic Hall algebra \cite{KS1},
the cohomological Hall algebra \cite{KSCOHA}, and its various ramifications, {\it e.g.} cluster algebras.
In section \ref{sec:Hecke} we will also discuss algebras of line operators localized on a surface operator
that preserve the same amount of symmetry and supersymmetry as BPS states discussed here.
In fact, line operators can be viewed as infinite mass limits of BPS states discussed here; this viewpoint explains
many similarities between algebras of BPS states discussed here and algebras of line operators discussed in section \ref{sec:lines}.

In the last line of \eqref{clcl} we illustrate the process of bound formation in the context of geometric engineering \eqref{surfgeom},
where from the Calabi-Yau viewpoint each BPS state in the bulk gauge theory is represented by a closed membrane on $\Sigma_g \subset CY_3$.
Similarly, BPS states localized
on a surface operator in the system \eqref{surfgeom} correspond to {\it open} membranes with boundary on $L$:
\bea
\CH_{\text{BPS}}^{\text{bulk}} & = & \CH_{\text{BPS}}^{\text{closed}} \quad (= \text{refined closed BPS states}) \\
\CH_{\text{BPS}}^{\text{surface}} & = & \CH_{\text{BPS}}^{\text{open}} \quad (= \text{refined open BPS states}) \nonumber
\eea
Specifically, the BPS states discussed here are, in fact, the so-called {\it refined} BPS states: besides grading by the charge
lattice $\Gamma$, their space has an additional $\Z$-grading by the difference between $U(1)_{23}$ and $U(1)_R$ symmetries.
{}From the viewpoint of a surface operator, this symmetry behaves in many ways as non-R flavor symmetry
and plays an important role in~\cite{Gukov:2004hz,Dimofte:2010tz}.
We will return to the role of this symmetry in section~\ref{sec:index}.

By analogy with \eqref{clcl}, when a bulk BPS state $\CB_1^{\text{bulk}} \in \CH_{\text{BPS}}^{\text{bulk}}$
forms a bound state with a BPS state localized on a surface operator $\CB_2^{\text{surface}} \in \CH_{\text{BPS}}^{\text{surface}}$
we obtain another BPS state localized on a surface operator $\CB_{12}^{\text{surface}} \in \CH_{\text{BPS}}^{\text{surface}}$:
\be
(\; \CB_1^{\text{bulk}} \;, \;\; \CB_2^{\text{surface}} \;) \quad \mapsto \quad \CB_{12}^{\text{surface}}
\label{opclosed}
\ee
$$
{\,\raisebox{-.5cm}{\includegraphics[width=5cm]{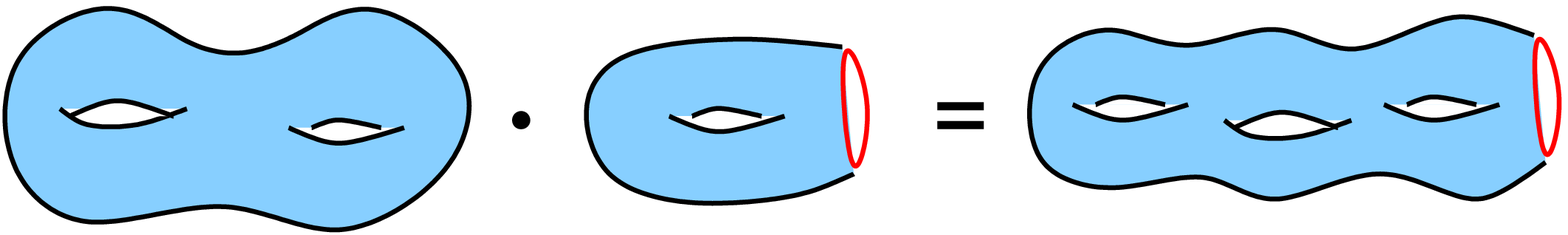}}\,}
$$
This defines an action of the algebra of bulk BPS states on the space of BPS states localized on a surface operator.
For example, when $CY_3$ is the total space of the $\CO (-1) \oplus \CO (-1)$ bundle over $\cp^1$ and $L$ is defined
by a knot in $S^3$ ({\it cf.} section~\ref{sec:3d3d}), the space of BPS states localized on a surface operator can be identified
with a homological knot invariant,
\be
\CH_{\text{BPS}}^{\text{surface}} \; \cong \; \CH_{\text{knot}}
\ee
and the action \eqref{HHrep} defines a plethora of anti-commuting operators ({\it i.e.} differentials)
acting on this space.

\subsection{Relation to 3d-3d correspondence and integrable systems}
\label{sec:3d3d}

The fivebrane configuration \eqref{surfgeom} encountered in the previous section has several interesting interpretations.
We already discussed the five-dimensional point of view: in gauge theory on $\R \times M$ the fivebrane defines a codimension-2
defect supported on $\R \times D$. Likewise, from the vantage point of $CY_3$ it defines a defect supported on a special
Lagrangian submanifold $L$ and relates BPS state count to enumerative invariants of the pair $(CY_3,L)$.

Here, we briefly comment on another interpretation of the system \eqref{surfgeom}, from the viewpoint of the fivebrane
observer on $R \times D$. It leads to yet another, equivalent description of physics --- including the spectrum of BPS
objects localized on a surface operator --- in terms of 3d $\CN=2$ theory that in general depends on both $L$ and $CY_3$.
Particular choices of the Calabi-Yau 3-fold that have been extensively studied in the literature and play an important role
in many applications include $CY_3 \cong \C^3$, $T^* L$, and the conifold geometry.
In particular, since neighborhood of any special Lagrangian submanifold $L$ looks like the total space of the cotangent bundle,
the choice $CY_3 \cong T^* L$ is especially canonical and depends only on $L$. In this case, the effective
3d $\CN=2$ theory on $\R \times D$ also depends only on the 3-manifold $L$ (and the total number of M5-branes),
so that we get a correspondence
\be
L \quad \leadsto \quad T[L] \,,
\label{3d3dbasic}
\ee
often called 3d-3d correspondence. In our presentation, we tried to emphasize its similarity to the study of surface operators.
Indeed, compactification of the system \eqref{surfgeom} on a circle, obtained by replacing $\R$ with $S^1$,
yields the familiar construction \eqref{surfeng} of a half-BPS surface operator in 4d gauge theory.
Moreover, there are many parallels between the space of SUSY vacua in the theory $T[L]$ on a circle
and the space of vacua in the surface operator theory. Both are described by algebraic equations
\be
\CM_{\text{SUSY}} \; = \; \{ A_i = 0 \}
\ee
which play the role of Ward identities for line operators in 3d $\CN=2$ theory \cite{DGG}.
Specifically, for 3d $\CN=2$ theories \eqref{3d3dbasic} labeled by 3-manifolds, the algebraic relations $A_i = 0$
define the moduli space of complex flat connections on $L$:
\be
\CM_{\text{SUSY}} (T[L]) \; = \; \CM_{\text{flat}} (L)
\label{mspaces}
\ee
Besides this basic property, there are many other elements of the dictionary between 3-manifolds
and 3d $\CN=2$ gauge theories that are described in \cite{DGG} and summarized in a companion contribution to this volume \cite{Dimofte-rev}.

As we reviewed in section~\ref{sec:AGT}, for applications to the AGT correspondence one is interested in turning on
the $\Omega$-background, so that $M = \R^4_{\epsilon_1,\epsilon_2}$ and $D = \R^2_{\epsilon_1}$.
This has the following effect on the surface operator theory or 3d theory $T[L]$,
where the role of $\epsilon_1$ and $\epsilon_2$ is clearly very different.
The $\Omega$-deformation {\it along} the surface operator controlled by the parameter $\epsilon_1$
has the effect of ``quantizing'' the system, {\it i.e.} replacing the polynomials $A_i$ by their non-commutative deformation
\be
A_i
~~\xrightarrow{~\epsilon_1 \ne 0~}~~
\widehat A_i
\ee
so that classical equations $A_i=0$ are replaced by the Schrodinger-like equations $\widehat A_i Z = 0$.
In a particular class of models where $A(x,y)=0$ realize spectral curves of integrable systems,
deformation by $\epsilon_1$ leads to Baxter equations of the corresponding integrable systems~\cite{Gadde:2013wq}.
For general values of the $S^1$ radius, the integrable systems in question are {\it trigonometric}
(also called {\it hyperbolic} in some of the literature), whose prominent examples include
the XXZ spin chain and the trigonometric Ruijsenaars model.

The role of $\epsilon_2$ is very different. Turning on $\epsilon_2 \ne 0$ (while keeping $\epsilon_1 = 0$)
does not make $A_i$ non-commutative and leads to the Nekrasov-Shatashvili duality~\cite{NS}
between $\CN=2$ theory on $S^1 \times D$ and, in general, a different integrable system.
The relation between the two integrable systems is some sort of spectral duality~\cite{Gadde:2013wq} which in the present
physical setup clearly corresponds to exchanging the role of $\epsilon_1$ and $\epsilon_2$
(or, equivalently, the support of surface operator inside $M = \R^4_{\epsilon_1,\epsilon_2}$).
Note, the two relations with integrable systems invoke rather different aspects, {\it e.g.} one goes via
Baxter equation, as was mentioned earlier, while the Nekrasov-Shatashvili correspondence goes via Bethe equations.
Conversely, Bethe equations are not manifest in a duality with $\epsilon_2 = 0$, while Baxter equations
are not manifest in a duality with $\epsilon_1 =0$.


\section{Surface operators and line operators}
\label{sec:lines}

Line operators remain, even at present time, the most familiar and better understood representatives
in the list of non-local operators in section~\ref{sec:what}.
They have a wide range of applications, from supersymmetric gauge theories --- where they play an important
role in computations of partition functions {\it a la} \cite{Pestun:2007rz,Gomis:2011pf,Gang:2012yr} as discussed
{\it e.g.} in a companion contribution to this volume \cite{Okuda-rev} ---
to phase structure of non-supersymmetric gauge theories, where they serve as excellent order parameters ({\it cf.} section~\ref{sec:order}).
This justifies the study of line operators in their own right.

Here, we will focus on rather specific aspects of line operators that have to do with how they interact with surface operators.
In fact, the main aspect we wish to discuss is that, in the presence of a surface operator, the OPE algebra of line operators becomes
non-commutative. And, then, we shall give some examples of such non-commutative structure and explain its simple geometric interpretation.
It is useful to keep in mind the parallel discussion of BPS states confined to a surface operator in section~\ref{sec:BPS}:
in both cases we deal with one-dimensional world-lines within a surface operator that lead to a non-commutative algebra.
Moreover, the rotation symmetry in two space-time dimensions transverse to the surface operator gives rise to a deformation
of the algebra by the parameter $q = e^{\hbar}$.

First, let us consider a four-dimensional gauge theory on a space-time manifold $M$ without surface operators.
(As usual, for concreteness, one can keep in mind a simple example of $M = \R^4$.)
It is well known that line operators form an algebra --- very similar to the algebra of BPS states discussed earlier ---
with the product
\be
L_1 \times L_2 \; \sim \; \sum_i V_i L_i
\label{LLproduct}
\ee
given by the operator product expansion (OPE).
In many familiar examples, that include topological and supersymmetric theories,
this product is commutative
simply because one can continuously exchange positions of line operators
by moving them around each other in four dimensional space.

One important aspect of the product \eqref{LLproduct} is that its coefficients $V_i$ are, in fact, vector spaces.
This aspect is not yet about surface operators {\it per se}, but does become more pronounced in the presence of surface operators,
as we explain shortly.
In many applications, $V_i$'s can be replaced by numbers, especially in situations where only dimensions
$v_i = \dim V_i$ are relevant to a particular application in question. This happens, for instance,
when line operators are compactified (either effectively or explicitly) on a circle.
Then, the OPE product of the resulting ``loop operators'' has the form of a typical OPE of local operators
\be
\CO_1 \times \CO_2 \; \sim \; \sum_i v_i \CO_i
\ee
with numerical coefficients $v_i$.

Now, let us imagine that line operators $L_1$ and $L_2$ are stuck to a two-dimensional subspace $D \subset M$,
as illustrate in Figure~\ref{vortex31}.
For example, as in our previous discussion, turning on the $\Omega$-background in directions transverse to $D$ inside $M$
will not allow line operators to move off from the surface $D$ without breaking the $SO(2)_{23}$ rotation symmetry
of the transverse space ($\cong \R^2_{\hbar}$).
What this means is the following:
The rotation symmetry $SO(2)_{23}$ makes $V_i$'s into graded vector spaces, graded by the angular momentum $h_{23}$.
And, therefore, the operator product expansion \eqref{LLproduct} has a ``refinement'' with graded vector spaces
as coefficients, as long as line operators are confined within the surface $D$.
In other words, the product \eqref{LLproduct} is commutative, but its graded version in general is not.

\FIGURE{
\includegraphics[width=2in]{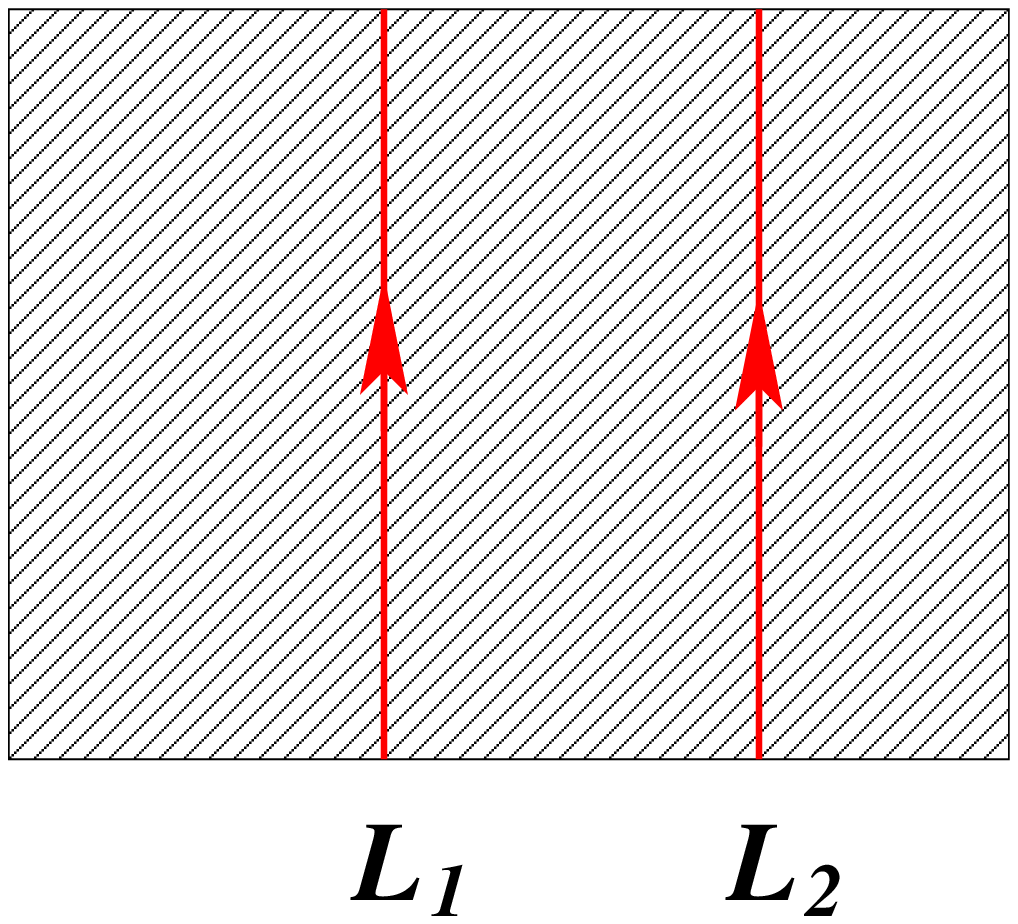}
\caption{Line operators confined to a surface operator do not commute.\\}
\label{vortex31}
}

A more dramatic way to make the product \eqref{LLproduct} non-commutative is to introduce surface operators supported on $D \subset M$.
This has several important ramifications.
First, it breaks the 4d Poincar\'e invariance and, therefore, does not allow to naively move around
line operators $L_1$ and $L_2$ in three transverse directions.
Moreover, in the presence of a surface operator there can exist additional line operators which are supported on the surface
operator and can not move into the rest of the 4-manifold $M$. Since such line operators are confined to the surface $D \subset M$,
they can not be passed through each other without encountering a singularity.
As a result, the OPE algebra of such line operators in general is non-commutative.

For example, in applications to AGT correspondence, one can consider line operators localized within
a surface operator in 4d $\CN=2$ theory $\CT [C, \frak{g}]$, where $C$ is a Riemann surface (possibly with punctures).
{}From the six-dimensional perspective reviewed in section \ref{sec:higher},
these are line operators localized on the 2-dimensional world-sheet $p \times D \times \{ 0 \}$
of a surface operator (= codimension-4 defect) in the six-dimensional $(2,0)$ theory on $C \times D \times \R^2_{\hbar}$,
where
\be
M = D \times \R^2_{\hbar}
\ee
is the 4d space-time of the $\CN=2$ gauge theory.
It was argued in \cite{Gukov:2006jk} that such line operators generate an affine Hecke algebra $H_{\text{aff}}$
of type $\frak{g}$ with parameter $q = e^{\hbar}$.
Note that this affine Hecke algebra is ``local on $C$.''
In other words, it does not depend on the details of the Riemann surface $C$ away from the point $p$.
This observation will be useful to us in what follows since for the purpose of deriving the non-commutative algebra
associated with the presence of surface operator (= ramification at $p \in C$) one can take $C$ to be something simple,
{\it e.g.} a torus or a disk.
For instance, in the basic case $G=SU(2)$ the affine Hecke algebra is
generated by $T$, $X$, and $X^{-1}$, which obey the relations (see {\it e.g.} \cite{ChrissG}):
\bea
&& (T+1)(T-q)=0 \nonumber \\
&& T X^{-1} - X T = (1-q) X \label{txxrels} \\
&& X X^{-1} = X^{-1} X = 1 \nonumber
\eea

The affine Hecke algebra has two close cousins: the affine Weyl group $\Weyl_{{\rm aff}}$,
which corresponds to the limit $q \to 1$,
and its ``categorification'', the affine braid group $B_{{\rm aff}}$.
They too have a simple physical interpretation which,
moreover, offers an intuitive explanation of the non-commutative product of line operators within a surface operator.
It follows from a compactification of our 2d-4d system on a circle, which can be achieved {\it e.g.} by taking
$D = S^1 \times \R$. From the six-dimensional perspective, reduction on $S^1$ gives the maximally supersymmetric Yang-Mills,
and a further reduction on $C$ yields a 3d $\CN=4$ sigma-model with target space $\CM_H (G,C)$, the moduli space
of Higgs bundles on $C$ (also known as the `Hitchin moduli space') \cite{Harvey:1995tg,Bershadsky:1995vm}.
The presence of a surface operator introduces ramification at $p \in C$, so that in the present case $\CM_H$
is the moduli space of ramified Higgs bundles~\cite{Gukov:2006jk}.

In the sigma-model on $\CM_H$, line operators correspond to functors acting on branes (or, boundary conditions).
According to (\ref{fundparam}), these functors form a group which often can be identified with the fundamental group
of the (sub)space of parameters of a surface operator.
Indeed, kinks on a surface operators are nothing but line operators.
In order to see this, consider a kink corresponding to an (adiabatic) variation of the continuous parameters of
a surface operator. On the one hand, it traverses a closed loop in the space of parameters. On the other hand,
it is localized in one dimension (= ``space'') and extended along the other dimension (= ``time'') on $D$,
just like line operators illustrated in Figure~\ref{vortex31}.

Line operators preserving certain symmetry and supersymmetry correspond to varying particular parameters
(that don't break these symmetries). For example, in applications to the geometric Langlands correspondence\cite{Kapustin:2006pk},
the Galois side corresponds to the $B$-model of $\CM_H$. Branes and boundary conditions that preserve this
particular supersymmetry are described by the derived category of coherent sheaves on $\CM_H$
and their charges are described by the K-theory.
Therefore, depending on whether one is interested in D-branes
(as objects of the derived category of coherent sheaves on $\CM_H$)
or in D-brane charges (classified by K-theory) one finds
the following groups acting the on the K-theory/derived category
of the moduli space of ramified Higgs bundles\footnote{For simplicity,
here we consider only one ramification point $p \in C$.
For the case of ramification at several points,
one finds several group actions, one for each ramification point.}:

\medskip
\noindent
{\bf Claim \cite{Gukov:2006jk}:}~~{\it affine Weyl group $\Weyl_{{\rm aff}}$ acts on $K(\CM_H)$}

~~~~~~~~~~~{\it affine Hecke algebra $H_{{\rm aff}}$ acts on $K^{\mathbb{C}^*} (\CM_H)$}

~~~~~~~~~~~{\it affine braid group $B_{{\rm aff}}$ acts on $D^b (\CM_H)$}

\noindent
This result can be regarded as a categorification of the affine
Hecke algebra, which in the local version of $\CM_H$ was also obtained
by Bezrukavnikov \cite{Bezrukavnikov} using a ``noncommutative
counterpart'' of the Springer resolution $\widetilde{\CN} \to \CN$.
The action of $\Weyl_{{\rm aff}}$ and $B_{{\rm aff}}$ in the first and the last
part of this claim can be understood as the monodromy action
in the space of parameters of the surface operator~\eqref{fundparam}.

For example, let us illustrate how this group action arises at
the level of D-brane charges, which are classified by $K(\CM_H)$.
The space of D-brane charges $K(\CM_H)$ varies as the fiber of
a flat bundle over the space of parameters
away from the points where $\CM_H$ develops singularities.
Since for the purposes of this question we are interested
only in the geometry of $\CM_H$, we can ignore the ``quantum''
parameter $\eta$. Hence, the relevant parameters are $(\alpha,\beta,\gamma)$,
which take values in the space, {\it cf.} (\ref{surfparameters}):
\begin{equation}
(\alpha,\beta,\gamma)
\in \big( \frak t \times \frak t \times \frak t \big) / \Weyl_{{\rm aff}}
\end{equation}
Moreover, $\CM_H$ becomes singular precisely for those values of
$(\alpha,\beta,\gamma)$ which are fixed by some element of $\Weyl_{{\rm aff}}$.
The set of such points is at least of codimension three in $\frak t^3$
(since it takes three separate conditions to be satisfied for
$(\alpha,\beta,\gamma)$ to be fixed by some element of $\Weyl_{{\rm aff}}$).
Therefore, the space of regular values of $(\alpha,\beta,\gamma) \in \frak t^3$
where $\CM_H$ is non-singular is connected and simply-connected,
and since $\Weyl_{{\rm aff}}$ acts freely on this space, the fundamental
group of the quotient is
\begin{equation}
\pi_1 \big( \{ (\alpha,\beta,\gamma) \}^{{\rm reg}} \big) = \Weyl_{{\rm aff}}
\end{equation}
This is the group that acts on D-brane charges, that is on $K(\CM_H)$.
In a similar way, one can deduce the action of the affine braid
group $B_{{\rm aff}}$ on $D^b (\CM_H)$ as the fundamental group of
the K\"ahler moduli space.
Indeed, for the $B$-model in complex structure $J$
the complexified K\"ahler parameters are $\eta + i \beta$,
and from (\ref{surfparameters}) one finds:
\begin{equation}
\pi_1 \big( \{ (\beta, \eta) \}^{{\rm reg}} \big) = B_{{\rm aff}}
\label{Baffquot}
\end{equation}
The same results can be derived more directly in the description of surface operators
as 2d-4d coupled systems.

\subsection{Line operators and Hecke algebras}
\label{sec:Hecke}

In section \ref{sec:what} we explained that a surface operator in 4d gauge theory can be equivalently
defined as a 2d sigma-model supported on $D$ and with a target space $X$ that has $G$-action.
Here we use this description of surface operators to explain more directly how algebra of line operators
localized on a surface operator leads to the affine Hecke algebra $H_{\text{aff}}$ or its close cousins
$\Weyl_{{\rm aff}}$ and $B_{{\rm aff}}$.
This particular approach offers an alternative derivation of the results in \cite{Gukov:2006jk} and to the best of
our knowledge has not appeared in the literature.
For concreteness and in order to keep things simple we shall restrict our attention to gauge theory with $G=SU(2)$.

When the surface operator is described by a sigma-model with target space $X$,
line operators (which act on branes on $X$) in turn can be viewed as branes on $X \times X$.
In the language of derived category, this means that an object $\CZ \in D^b (X \times X)$
called the ``kernel'' defines an exact functor 
$\Phi_{\CZ}: D^b (X) \to D^b (X)$, such that
\be
\label{kernfunctor}
\Phi_{\CZ} (\CE) := p_{2*} (\CZ \otimes p^*_1 (\CE))
\ee
where $p_1$ (resp. $p_2$) is the projection to the first (resp. the second)
factor.\footnote{To be more precise, the pull-back $p_1^*$ is left-derived and
the push-forward $p_{2*}$ is right-derived.}
We can define a product of two line operators $\CA$ and $\CB$
by bringing them together (as in Figure~\ref{vortex31}) which leads
to a composition of the transforms $\Phi_{\CA}$ and $\Phi_{\CB}$.
This gives a new transform
\be
\label{abfunctor}
\Phi_{\CB \star \CA} \cong \Phi_{\CA} \circ \Phi_{\CA}
\ee
with the kernel
\be
\label{kcompos}
\CB \star \CA = p_{13*} (p_{12}^* \CA \otimes p_{23}^* \CB)
\ee
where $p_{ij}$ are the obvious projection maps from
$X \times X \times X$ to $X \times X$.
In particular, the diagonal $\Delta_X: X \hookrightarrow X \times X$ gives the identity.
The product \eqref{kcompos} is associative
\be
\label{abcabc}
\CC \star (\CB \star \CA) \cong (\CC \star \CB) \star \CA
\ee

We are interested in the case where $X = \CN$ is the nilpotent cone
for $SL(2,\C)$ or its Springer resolution $\tilde \CN$.
This is a special case of a larger class of examples where $X$
is (the minimal resolution) of  of the Kleinian quotient singularity
$\C^2/\Gamma$ for a finite subgroup $\Gamma \subset SL(2,\C)$.
In this case, there is an equivalence (the derived McKay correspondence):
\be
\label{dmckay}
D^b (X) \cong D^b_{\Gamma} (\C^2)
\ee
The category $D^b_{\Gamma} (\C^2)$ has simple objects
\be
\label{sbranes}
S_i = \rho_i \otimes \CO_p
\ee
where $\rho_i$ are irreducible representations of $\Gamma$
and $\CO_p$ is the skyscraper sheaf supported at the origin of $\C^2$.
These are precisely the fractional branes on $\C^2/\Gamma$.
In the derived category of the minimal resolution $X$,
the  simple objects \eqref{sbranes} are represented by \cite{KVesserot},
\bea
S_0 & = & \CO_{\sum C_i} \label{sbrresol} \\
S_i & = & \CO_{C_i} (-1)[1] \nonumber
\eea
where $C_i$ are the exceptional divisors.

An important feature of fractional branes is that,
in the derived category of $X$, they are described by
spherical objects and, therefore, according to the results
of Seidel and Thomas \cite{SThomas}, define twist functors $T_i$
which generate the braid group $Br (\Gamma)$.
As the name suggests, an object $\CE \in D^b (X)$ is called
$d$-spherical if ${\rm Ext}^* (\CE , \CE)$ is isomorphic to
$H^* (S^d,\C)$ for some $d>0$,
\be
\label{sphericale}
{\rm Ext}^i (\CE , \CE) =
\begin{cases}
\C & \text{if } i=0 \text{ or } d \\
 0 & \text{otherwise}
\end{cases}
\ee
A spherical B-brane defines a twist functor
$T_{\CE} \in {\rm Auteq} (D^b (X))$
which, for any $\CF \in D^b (X)$, fits into exact triangle
\be
\label{twisttriangle}
{\rm Hom}^* (\CE , \CF) \otimes \CE \longrightarrow \CF \longrightarrow T_{\CE} (\CF)
\ee
where the first map is evaluation.
The functor $T_{\CE}$ can be written as a Fourier-Mukai transform \eqref{kernfunctor}
associated with the brane $\CZ$ on $X \times X$,
\be
\label{stcone}
\CZ = {\rm Cone} \left( \CE^{\vee} \btimes \CE \to \CO_{\Delta_X} \right)
\ee
where
$\CE^{\vee}$ denotes the dual complex, $\Delta_X$ is the diagonal in $X \times X$,
and $\CE \btimes \CF = p_2^* \CE \otimes p_1^* \CF$ is the exterior tensor product.
At the level of cohomology, the twist functor $T_{\CE}$ acts as,
$$
x \mapsto x + (v (\CE) \cdot x) ~v (\CE)
$$
where $v(\CE) = ch (\CE) \sqrt{Td (X)} \in H^* (X)$
is the D-brane charge (the Mukai vector) of $\CE$.
Summarizing, ``spherical branes'' (spherical objects in $D^b(X)$)
lead to autoequivalences of $D^b (X)$. What is the group they generate?

Given an $A_{n}$ chain of spherical objects, that is a collection
of spherical objects $\CE_1, \ldots, \CE_{n}$ which satisfy the condition
\be
\label{sphericalderived}
\sum_k \dim {\rm Ext}^k (\CE_i , \CE_j)
\begin{cases}
1 & |i-j|=1 \\
0 & |i-j| >1
\end{cases}
\ee
with some minor technical assumptions Seidel and Thomas \cite{SThomas}
showed that the corresponding twist functors $T_{\CE_i}$
generate an action of the braid group $Br_{n+1}$ on $D^b (X)$.
More generally, a chain of spherical objects associated with $\Gamma$
gives rise to the action of the braid group $Br(\Gamma)$ on B-branes.
The generators of $Br (\Gamma)$ correspond to vertices
of the affine Dynkin diagram of $\Gamma$ and obey the relations
\be
\label{braidrels}
T_i T_j T_i = T_j T_i T_j
\ee
if the vertices $i$ and $j$ are connected by an edge, and $T_i T_j = T_j T_i$ otherwise.
In particular, in the situation we are interested in,
namely when $X$ is the minimal resolution quotient singularity $\C^2/\Gamma$,
the braid group $Br (\Gamma)$ is the essential part of
the group of autoequivalences of $X$.
Specifically, the group of autoequivalences
of $X$ is \cite{IshiiUehara,Bridgelandii}:
\be
\label{cnauteq}
{\rm Auteq} (D^b(X)) = \Z \times \left( {\rm Aut} (\Gamma) \ltimes Br (\Gamma) \right)
\ee
where the first factor is generated by the shift functor $[1]$,
the group ${\rm Aut} (\Gamma)$ is the group of symmetries of
the affine Dynkin diagram associated to $\Gamma$,
and $Br(\Gamma)$ is the braid considered above.

In particular, for $\Gamma = \Z_2$ which is relevant to the $SU(2)$ gauge theory, the group ${\rm Auteq} (D^b(X))$
is generated by the functors $T_{\pm}$ and $R$ (and the shift functor, of course).
Indeed, in the present case there are two spherical objects (two fractional branes \eqref{sbranes})
which lead to the twist functors $T_+$ and $T_-$.
These two are exchanged by $R$, the generator of ${\rm Aut} (\Gamma) \cong \Z_2$,
so that in total we obtain the group generated by $T_{\pm}$ and $R$ which obey the relations
\bea
T_+ R & = & R T_- \label{sutwoauteq} \\
R^2 & = & 1 \nonumber
\eea
Notice, for $\Gamma = \Z_2$ there are no braid relations of the form \eqref{braidrels}.
Nevertheless, we still shall refer to the resulting group as the affine braid group of type $\hat A_1$.

To make contact with our earlier discussion, we can identify the autoequivalences that generate the braid group
with the monodromies in the category of B-branes around special points in the K\"ahler moduli space of $X$.
In the case at hand, there are three such points: $i)$ the large volume limit, $ii)$ the ``conifold limit''
(where $X = \C^2 / \Z_2$ with zero $B$-field),
and $iii)$ the orbifold limit (where $X = \C^2 / \Z_2$ with $B=\frac{1}{2}$).
A monodromy around each of these points defines a Fourier-Mukai transform
associated with a certain brane on $X \times X$. Following \cite{Aspinwallrev},
we denote these branes, respectively, as $\CL$, $\CK$, and $\CG$.
The corresponding transforms will be denoted by $\Phi_{\CL}$, $\Phi_{\CK}$, and $\Phi_{\CG}$.

As we shall see below, the monodromy $\Phi_{\CL}$ has infinite order and,
therefore, is related to the generator $X$ in \eqref{txxrels}
or the generator $T_{\pm}$ in \eqref{sutwoauteq}.
On the other hand, while the monodromies around the conifold point
and the orbifold point are both of order 2 at the level of K-theory charges,
in the derived category we have
\be
\label{tktg}
\Phi_{\CG}^2 = 1 \quad,\quad \Phi_{\CK}^2 \ne 1
\ee
Therefore, $\Phi_{\CG}$ which comes from the quantum $\Z_2$ symmetry of $\C^2 / \Z_2$
should be identified with the generator $R$ in \eqref{sutwoauteq} which has a similar origin.
The monodromy around the orbifold point is a composition of the monodromies
around the conifold point and the large radius limit. Therefore, from \eqref{abfunctor} we get
\be
\label{gislk}
\CG = \CK \star \CL
\ee

Now, let us describe explicitly $\CL$, $\CK$, and $\CG$, and verify \eqref{tktg}.
The monodromy around the large radius limit is always associated with
\be
\label{lrlkernel}
\CL = \CO_{\Delta_X} (1)
\ee
Since D-brane wrapped on the exceptional divisor $C$ becomes massless at the conifold point,
the monodromy around the conifold point is the twist functor $T_{\CE}$ associated with the spherical object $\CE = \CO_{C}$.
Therefore,
\be
\label{kkernel}
\CK = \left( \CO_C^{\vee} \btimes \CO_C \to \CO_{\Delta_X} \right)
\ee
and using \eqref{gislk} we get
\be
\label{gkernel}
\CG = \left( \CO_C (-1)^{\vee} \btimes \CO_C \to \CO_{\Delta_X} (1) \right)
\ee
Now, to verify \eqref{tktg} we can either compute how the functors $\Phi_{\CL}$, $\Phi_{\CK}$,
and $\Phi_{\CG}$ act on simple branes, such as the 0-brane $\CO_p$, or to study their
composition using \eqref{abfunctor}. In particular, computing $\Phi_{\CK}^n (\CO_p)$ we can
verify that $\Phi_{\CK}$ is indeed of infinite order.
Similarly, we find
$$
\CG \star \CG = \CO_{\Delta_X}
$$
which is the first relation in \eqref{tktg}.

The transforms $\Phi_{\CL}$, $\Phi_{\CK}$, and $\Phi_{\CG}$ are autoequivalences of $D^b (X)$.
In fact, they generate the entire group \eqref{sutwoauteq} which we found
earlier by looking at the fractional branes on $\C^2 / \Z_2$.
This can be shown by explicitly matching the generators.
First, one of the generators $T_{\pm}$ is the twist functor
associated with the spherical object $\CE = \CO_C$.
Without loss of generality, we assume that this generator is $T_+$.
According to \eqref{kkernel}, it should be identified with the monodromy
around the conifold point, $\Phi_{\CK}$.
Similarly, the order-2 generator $R$ should be identified with
the monodromy around the orbifold point $\Phi_{\CG}$ which is
also of order 2 and has a similar origin (both come from the quantum $\Z_2$ symmetry of $\C^2 / \Z_2$).
Summarizing,
\bea
& & \Phi_{\CK} \longleftrightarrow T_+^{-1} \label{rtmatch} \\
& & \Phi_{\CG} \longleftrightarrow R \nonumber
\eea
The remaining generator can be expressed as a product of these two, {\it cf.} \eqref{sutwoauteq} and \eqref{gislk}.
In particular, we conclude that the monodromy
around the large volume limit $\Phi_{\CL}$ should be identified with $T_+ R = R T_-$.

As we mentioned earlier, at the level of K-theory charges the OPE algebra
of line operators that we are considering should reduce to the affine
Weyl group $\Weyl_{aff}$ or affine Hecke algebra $H_{aff}$.
{}From \eqref{lrlkernel} -- \eqref{gkernel} it is easy to see that
the monodromies $\Phi_{\CL}$, $\Phi_{\CK}$, and $\Phi_{\CG}$
act on the charges of D0, D2, and D4 branes as
\be
\label{mmmonodromies}
M_{\CL} =
\begin{pmatrix}
1 & 0 & 0 \\
1 & 1 & 0 \\
\frac{1}{2} & 1 & 1
\end{pmatrix}
\quad,\quad
M_{\CK} =
\begin{pmatrix}
1 & 0 & 0 \\
0 & -1 & 0 \\
0 & 0 & 1
\end{pmatrix}
\quad,\quad
M_{\CG} =
\begin{pmatrix}
1 & 0 & 0 \\
-1 & -1 & 0 \\
\frac{1}{2} & 1 & 1
\end{pmatrix}
\ee
It is easy to verify that
\be
\label{mgmkml}
M_{\CG} = M_{\CK} M_{\CL}
\ee
in agreement with \eqref{gislk},
and that both $M_{\CK}$ and $M_{\CG}$ are of order two:
\be
\label{mkmg}
M_{\CK}^2 = 1 \quad\quad,\quad\quad M_{\CG}^2 = 1
\ee
Via the identification \eqref{rtmatch} this implies that, in addition
to the relation $R^2=1$ which is already included in \eqref{sutwoauteq},
we need to impose an extra condition $T_+^2=1$ which, of course, implies $T_-^2=1$ as well:
\be
\label{ttone}
T_i^2 = 1
\ee
Therefore, at the level of K-theory charges, the group generated by
the monodromies is a semidirect product of $\Z_2$ (generated by $T_+$) and $\Z$ (generated by $T_+ R$).
This is precisely the affine Weyl group $\Weyl_{\text{aff}}$ for $G=SU(2)$.

More generally, instead of the quadratic relations \eqref{ttone},
we can consider imposing extra relations
\be
\label{thecke}
T_i^2 = (q^{1/2} - q^{-1/2}) T_i + 1
\ee
on all the generators $T_i \in B_{\text{aff}}$ which correspond to the simple reflections in $\Weyl_{\text{aff}}$.
As we discussed earlier, this should lead to the affine Hecke algebra $H_{\text{aff}}$.
In the example we are considering, we require $T_+$ (and $T_-$) to obey \eqref{thecke}.
Furthermore, motivated by the specialization to
the affine Weyl group considered above, we introduce the notations
\bea
T & = & q^{1/2} T_+ \label{txviattr} \\
X & = & T_+ R = R T_- \nonumber
\eea
Then, the quadratic constraint \eqref{thecke} on $T_+$ implies a similar constraint on $T$,
\be
\label{thaffsutwo}
T^2 = (q-1) T + q
\ee
which is precisely one of the relations in the affine Hecke algebra \eqref{txxrels}.
Moreover, from \eqref{thecke} we obtain
$$
T_{\pm}^{-1} = T_{\pm} + (q^{-1/2} - q^{1/2})
$$
which can be used to find
\be
\label{xsutwoinverse}
X^{-1} = T_- R + (q^{-1/2} - q^{1/2}) R
\ee
Now, using \eqref{sutwoauteq}, \eqref{txviattr}, and \eqref{xsutwoinverse},
it is easy to check that $X$ and $T$ satisfy
\be
\label{xtsutwo}
T X^{-1} - X T = (1-q) X
\ee
which is precisely the second relation in the affine Hecke algebra \eqref{txxrels}.
Therefore, we verified that imposing extra relations \eqref{thecke}
on the generators of the affine braid group leads to the affine Hecke algebra.

Conversely, starting with \eqref{thecke} and using the identifications \eqref{txviattr},
it easy to verify that $T_{\pm}$ and $R$ obey \eqref{sutwoauteq} with
the additional relations \eqref{thecke}. For example,
\be
\label{rviaxt}
R = q^{-1/2} TX + (q^{-1/2} - q^{1/2}) X
\ee
so that after a little algebra we get $R^2 = 1$.

Finally, we conclude our discussion of line operators confined to a surface operator
with one more deformation of the Hecke algebra that depends on two deformation parameters, $q$ and $t$.
This deformation is called the {\it double affine Hecke algebra}, or DAHA for short.
It already made an appearance in the physical literature~\cite{Aganagic:2011sg,Cherednik:2011nr}
on refined BPS states and knot invariants
that we mentioned in section~\ref{sec:BPS} and cries out for an interpretation either
as algebra of BPS states or algebra of line operators.

In fact, a convenient starting point for defining DAHA (which for simplicity we explain in the basic case of $G = SU(2)$)
is the orbifold fundamental group of the elliptic curve quotient, {\it cf.} \eqref{Baffquot}:
\be
\pi_1^{\text{orb}} ( \{ E \setminus 0 \} / \Z_2 )
\cong
\pi_1^{\text{orb}} ( \{ E \times E \setminus \text{diag} \} / \Z_2 )
\label{Bellorb}
\ee
generated by $X$, $Y$, and $T$ with the relations
\bea
&& TXT = X^{-1} \nonumber \\
&& TY^{-1}T = Y \label{orbEE} \\
&& Y^{1} X^{-1} YX T^2 = 1 \nonumber
\eea
Deforming the last relation to $Y^{1} X^{-1} YX T^2 = t^{-1/2}$ gives the so-called {\it elliptic braid group} $B_{\text{ell}}$.
Furthermore, imposing by now familiar quadratic Hecke relation as in \eqref{txxrels}, \eqref{thecke}, or \eqref{thaffsutwo}
with another deformation parameter $q$ leads to the complete definition of DAHA:
\be
\text{DAHA} \; = \; \C [B_{\text{ell}}] / ((T - q^{1/2}) (T + q^{-1/2}))
\ee
as a quotient of the group algebra of $B_{\text{ell}}$.
The operator $Y$ in this algebra is called the difference Dunkl operator.

Comparing the fundamental group \eqref{Bellorb} to \eqref{fundparam}, \eqref{aetaabelian} and \eqref{surfparameters},
we see that it classifies monodromies of the parameters $(\alpha, \eta) \in [( \TT \times \TT^{\vee} ) / \Weyl]^{\text{reg}}$
of a surface operator in $SU(2)$ gauge theory.
In other words, it classifies line operators that correspond to monodromies of the parameters $\alpha$ and $\eta$,
which are precisely the parameters of the basic surface operator in $\CN=2$ gauge theory, {\it cf.} section \ref{sec:N2}.
Of course, such surface operators can be also embedded in $\CN=4$ gauge theory (with the same gauge group) where \eqref{Bellorb}
means classifying monodromies of $\alpha$ and $\eta$, while keeping $\beta$ and $\gamma$ fixed.
This is similar to what we encountered in \eqref{Baffquot}, except that now the relevant problem
involves the B-model of $\CM_H$ in complex structure $I$, where $\beta + i \gamma$ is the complex structure parameter,
while $\eta + i \alpha$ represents the K\"ahler modulus (stability condition).
See \cite{Gukov:2006jk} for more details.

Therefore, we expect that the algebra of line operators confined to a surface operator in $\CN=2$ gauge theory
(or in $\CN=4$ gauge theory with supersymmetry of type $B_I$)
is intimately related, if not equal, to the double affine Hecke algebra.
It would be interesting to tackle a similar physical realization of quantum affine algebras and Kac-Moody algebras
acting on the equivariant K-theory of certain quiver varieties constructed by Nakajima \cite{Nakajima1,Nakajima2}.
It is natural to expect that this action can be lifted to an action of the fundamental group of the K\"ahler moduli space
on the derived category of the quiver variety involved in this construction.


\section{Superconformal index}
\label{sec:index}

The AGT correspondence has a sister that, on the one hand, is simpler, but in another respect is more mysterious.
It relates another observable (partition function) of the 4d $\CN=2$ theory $\CT[C]$ to the partition function of
a non-supersymmetric 2d theory on the Riemann surface $C$,
\be
\CI_{4d} (\CT[C]) \; = \; Z_{2d} (C)
\label{indexvsZ2}
\ee
where $\CI_{4d} (\CT[C])$ is a superconformal index of the theory $\CT[C]$, defined as
\be
{\cal I}_{4d}(\fp,\fq,\ft)={\rm Tr} (-1)^F \fp^{h_{23}-r} \fq^{h_{01}-r} \ft^{R+r}.
\label{indexdef}
\ee
For a theory with weakly coupled Lagrangian description the index is computed by a matrix integral:
\be\label{4dcompute}
{\cal I}_{4d}(\fp,\fq,\ft)=\int [dU] \exp \Big( \sum_{n=1}^{\infty} \sum_{j}\frac{1}{n} f^{(j)_{4d}}(\fp^n,\fq^n,\ft^n) \chi_{R_j}(U^n,V^n)\Big).
\ee
Here, $U$ and $V$ denote elements of gauge and flavor groups, respectively. The invariant Haar measure integral $\int [dU]$ imposes the Gauss law over the Fock space. The sum is  over different ${\cal N}=2$ supermultiplets appearing in the Lagrangian,
with $R_j$ being the representation of the $j$-th multiplet under gauge and flavor group, and $\chi_{R_j}$ the character of $R_j$. The function $f^{(j)}$ is called single letter index. It is equal to either $f^{\text{vector}}_{4d}$ or $f^{\text{half-hyper}}_{4d}$ depending on whether the $j$-th multiplet is ${\cal N}=2$ vector multiplet or half-hypermultiplet \cite{Gadde:2011uv}:
\bea
f^{\text{vector}}_{4d}=\frac{-\fp-\fq-\ft+2\fp\fq+\fp\fq/\ft}{(1-\fp)(1-\fq)}
\qquad\quad f^{\text{half-hyper}}_{4d}= \frac{\sqrt{t}-\fp\fq/\sqrt{t}}{(1-\fp)(1-\fq)}.
\label{4dff}
\eea

\begin{table}
\begin{centering}
\begin{tabular}{|c|c|}
\hline
$4d$ & $2d$\tabularnewline
\hline
\hline
$\{\CQ_{-}^{1},(\CQ_{-}^{1})^{\dagger}\}=E+h_{01}+h_{23}-(2R+r)$ & broken \tabularnewline
\hline
$\{\CQ_{+}^{1},(\CQ_{+}^{1})^{\dagger}\}=E+h_{01}-h_{23}-(2R+r)$ & $\{\CG_{\ttL}^{+},(\CG_{\ttL}^{+})^{\dagger}\}=2H_{\ttL}- J_{\ttL}$\tabularnewline
\hline
$\{\CQ_{-}^{2},(\CQ_{-}^{2})^{\dagger}\}=E-h_{01}+h_{23}+(2R-r)$ & $\{\CG_{\ttR}^{+},(\CG_{\ttR}^{+})^{\dagger}\}=2H_{\ttR}+ J_{\ttR}$\tabularnewline
\hline
$\{\CQ_{+}^{2},(\CQ_{+}^{2})^{\dagger}\}=E+h_{01}-h_{23}+(2R-r)$ & broken \tabularnewline
\hline
$\{\widetilde{\CQ}_{\dot{-}}^{1},(\widetilde{\CQ}_{\dot{-}}^{1})^{\dagger}\}=E-h_{01}-h_{23}-(2R-r)$ & $\{\CG_{\ttR}^{-},(\CG_{\ttR}^{-})^{\dagger}\}=2H_{\ttR}- J_{\ttR}$\tabularnewline
\hline
$\{\widetilde{\CQ}_{\dot{+}}^{1},(\widetilde{\CQ}_{\dot{+}}^{1})^{\dagger}\}=E+h_{01}+h_{23}-(2R-r)$ & broken \tabularnewline
\hline
$\{\widetilde{\CQ}_{\dot{-}}^{2},(\widetilde{\CQ}_{\dot{-}}^{2})^{\dagger}\}=E-h_{01}-h_{23}+(2R+r)$ & broken \tabularnewline
\hline
$\{\widetilde{\CQ}_{\dot{+}}^{2},(\widetilde{\CQ}_{\dot{+}}^{2})^{\dagger}\}=E+h_{01}+h_{23}+(2R+r)$ & $\{\CG_{\ttL}^{-},(\CG_{\ttL}^{-})^{\dagger}\}=2H_{\ttL}+ J_{\ttL}$\tabularnewline
\hline
\end{tabular}
\par\end{centering}
\caption{\label{commutation}Embedding of $2d$ $\CN=(2,2)$ algebra into $4d$ $\CN=2$.}
\end{table}

Now let us incorporate half-BPS surface operators supported on $(x^0,x^1)$ plane in a four-dimensional space-time.
As discussed in section~\ref{sec:N2} ({\it cf.} Table~\ref{tab:2d4dtarget}), such surface operators preserve
$SU(1,1|1)_\ttL\times SU(1,1|1)_\ttR \times U(1)_e$ subgroup of the superconformal symmetry group,
which is basically the superconformal symmetry of a 2d $\CN=(2,2)$ theory on the $(x^0,x^1)$ plane.
The standard bosonic generators of this two-dimensional superconformal algebra can be easily identified via embedding
into 4d $\CN=2$ algebra summarized in Table~\ref{commutation}:
\bea
H_{\ttL,\ttR} & = & \frac{1}{2} \left( E \pm h_{01} \right) \,, \\
J_{\ttL,\ttR} & = & h_{23} + (2R \pm r) \nonumber
\eea
Luckily, the unbroken part of the symmetry and supersymmetry suffices for defining the superconformal index \eqref{indexdef}
even in the presence of surface operators, with all of the fugacities.
Moreover, since half-BPS surface operators can be defined via coupling to 2d $\CN=(2,2)$ theory supported on $D$,
it gives a very convenient way of computing the index: one simply needs to add the contribution of 2d degrees of freedom,
namely the so-called ``flavored elliptic genus'' of the 2d $\CN=(2,2)$ system \cite{Gadde:2013wq}:
\be
\label{2dindexdef}
\CI_{2d} (a_j;q,t) \; =\; {\rm Tr} (-1)^F q^{H_\ttL + \frac{1}{2} J_\ttL} t^{-J_\ttL} \prod_j a_j ^{f_j}.
\ee
As the ordinary elliptic genus \cite{Witten}, it depend on the ``Jacobi variables'' $q$ and $t$
that can be identified with the 4d fugacities $(\fp,\fq,\ft)$ by means of the embedding in Table~\ref{commutation}:
\be
q=\fq, \qquad t=\fp\fq/\ft, \qquad e=\fp^2/\ft.
\ee
Much like the basic building blocks of 4d $\CN=2$ theories are vector- and hyper-multiplets whose contributions
to the index are summarized in \eqref{4dff}, the basic building blocks of 2d $\CN=(2,2)$ theories are chiral and vector multiplets.
Their contributions to the flavored elliptic genus, respectively, are
\be
\CI_{2d \text{ chiral}} \; = \;
\frac{\theta(at;q)}{\theta(a;q)}
\ee
and
\be
\label{UNvector}
\CI_{2d \text{ vector}}^{U(n)} \; = \; \Big(\frac{(q;q)^{2}}{\theta(t;q)}\Big)^{n} \prod_{i\neq j}\Big((1-\frac{a_i}{a_j})
\frac{\theta(ta_i/a_j;q)}{\theta(a_i/a_j;q)} \Big)^{-1}
\ee
where $\theta(x;q):=(x;q)(q/x;q)$ and $(x;q)=\prod_{i=0}^{\infty}(1-xq^i)$.
With these basic tools one can easily compute the index of any 2d-4d coupled system
that, as in section~\ref{sec:what}, describes a fairly generic surface operator.
One interesting question, that still remains open since the pioneering work \cite{Gadde:2011uv},
is the identification of 2d TQFT whose partition function on a Riemann surface $C$ matches the index in \eqref{indexvsZ2}.

For further discussion of the superconformal index with surface operators see {\it e.g.}
\cite{Nakayama:2011pa,Gaiotto:2012xa,Iqbal:2012xm,Gadde:2013dda,Bullimore:2014nla}.


\section{Surface operators as order parameters}
\label{sec:order}

Finally, going back to the origins, we wish to explain that surface operators can serve as order parameters,
in particular, they can distinguish the IR phases of 4d gauge theories.
Typically, the information captured by surface operators is roughly equivalent to the spectrum of line operators
in the low-energy theory \cite{Gukov:2013zka}.

Extending standard arguments that show how 't Hooft and Wilson line operators exhibit ``area law'' in Higgs and confining phases,
respectively, one can quickly conclude that surface operators can exhibit a ``volume law'' in phases that admit domain walls which
can end on surface operators.
Not much is known about such peculiar domain walls, and exploring this direction would be an excellent research topic.
In particular, by studying the spectra of domain walls in 4d supersymmetric gauge theories one might hope to learn whether
surface operators can detect interesting phases not distinguished by Wilson and 't Hooft operators, {\it cf.} \cite{Cachazo:2002zk}.

Here, we present a slightly different mechanism for the ``volume law'' behavior due to thermal effects.
This material is new and has not appeared in the previous literature.
It will also give us an excellent opportunity to illustrate how surface operators are described in the holographic dual
of gauge theory, which is a convenient way to study thermal physics.
As usual, in order to study 4d gauge theory at finite temperature $T$,
we compactify the time direction on a circle of circumference $\beta = 2 \pi / T$
and study the theory on a space-time manifold $M = S^1_{\beta} \times S^3$
with thermal (anti-periodic) boundary conditions on fermions.
Following \cite{Witten:1998zw},
we can study this system using a holographic dual description,
which is available for many 4d gauge theories.

For concreteness, let us focus on
${\cal N}=2$ gauge theory with a massless adjoint hypermultiplet or, equivalently, ${\cal N}=4$ super-Yang-Mills,
for which the gravity dual is especially simple and well studied \cite{Maldacena:1997re,Witten:1998qj,Gubser:1998bc}.
After all, at finite temperature the precise details of the spectrum and interactions are expected to be less important
and we expect that results should apply to more general ${\cal N}=2$ gauge theories.
Surface operators in this theory are well understood
and have a simple description in the holographic dual \cite{Gukov:2006jk,Gukov:2008sn,Gomis:2007fi}.

Specifically, the holographic dual of ${\cal N}=4$ super-Yang-Mills
is type IIB string theory on $X_5 \times S^5$, where the 5-manifold $X_5$
is either a ``thermal AdS''
$$
X_5 \; \cong \; B^4 \times S^1 \qquad (\text{low temperature})
$$
in the low temperature phase, or the Schwarzschild black hole on AdS space
$$
X_5 \; \cong \; S^3 \times B^2 \qquad (\text{high temperature})
$$
in the high temperature phase \cite{Witten:1998zw}.
Note, that both of these manifolds are bounded by $S^1 \times S^3$,
which is precisely $M$ where the boundary gauge theory lives.

\FIGURE{
\includegraphics[height=1.5in,width=1.5in]{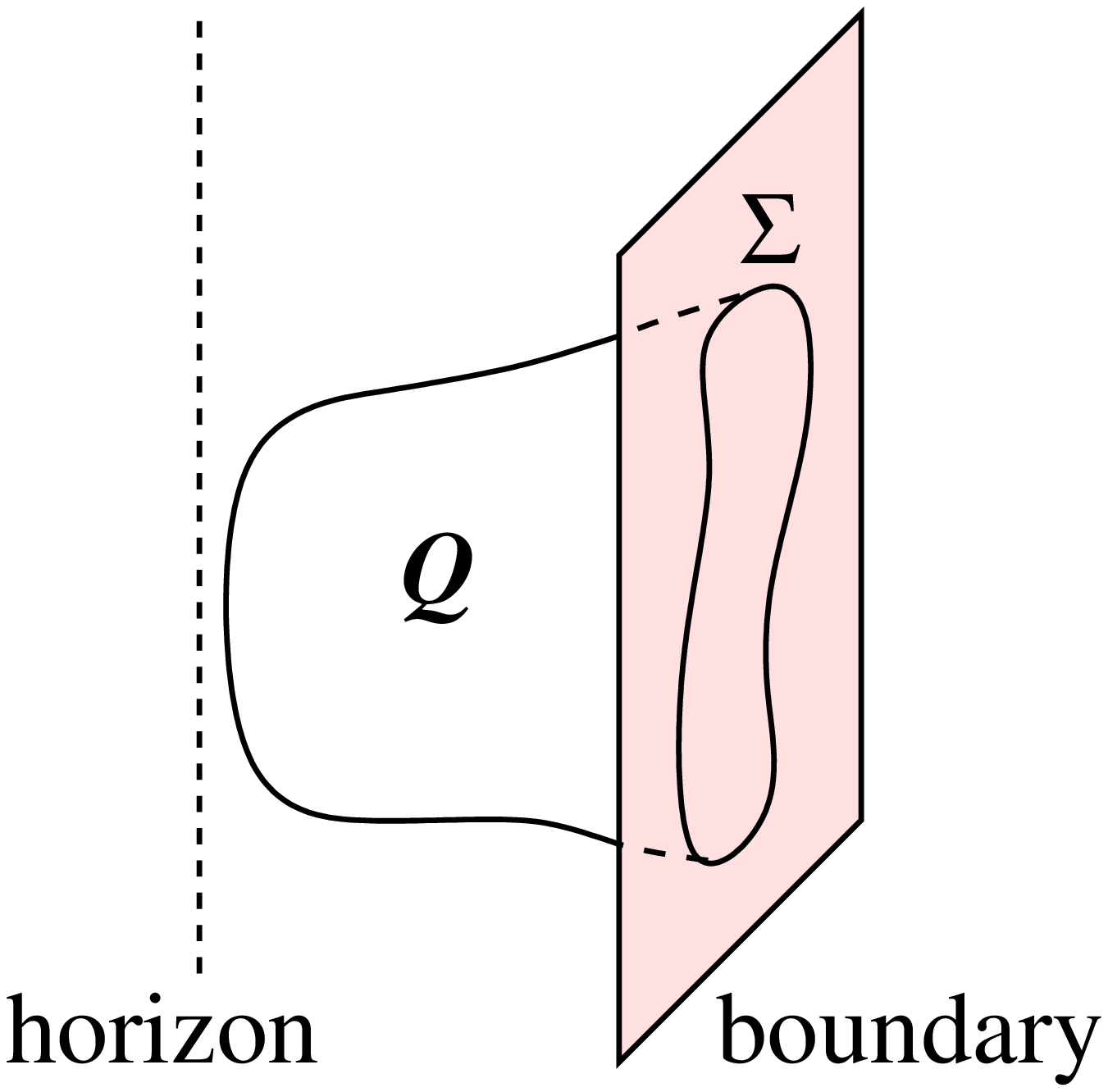}
\caption{A surface operator of the boundary theory
is represented by a D3-brane in the holographic dual.\\}
\label{amoeba}
}

Now, we can introduce surface operators supported on $D \subset M$.
For generic values of the continuous parameters $\alpha$ and $\eta$,
in the holographic dual such surface operators are represented by
D3-branes with four-dimensional world-volume
$$
Q \times S^1 \; \subset \; X_5 \times S^5
$$
where $Q \subset X_5$ is a volume-minimizing 3-dimensional submanifold
bounded by $D = \partial Q$, and $S^1$ is a great circle in the $S^5$.
Indeed, notice that such a D3-brane probe breaks the isometry / superconformal symmetry precisely as described in  Table \ref{tab:2d4dtarget}.

There are two qualitatively different choices of $D$,
which correspond to spatial surface operators with $D \subset S^3$
or temporal surface operators with $D = \gamma \times S^1_{\beta}$,
for some closed path $\gamma \subset S^3$.
In the low temperature confining phase we have
$$
\langle {\cal O}_{\text{temporal}} \rangle \; = \; 0
$$
since $S^1_{\beta}$ is not contractible in $X_5$,
and so there is no minimal submanifold $Q$ bounded by $D$.
On the other hand, spatial surface operators exhibit the area law in this phase:
$$
\langle {\cal O}_{\text{spatial}} \rangle \; \sim \; e^{-\text{Area}(D)}
$$

As we decrease the value of $\beta$, the theory undergoes a phase transition
to a deconfining phase \cite{Witten:1998zw} with $X_5 \cong S^3 \times B^2$,
which does admit minimal submanifolds $Q \cong \gamma \times B^2$ bounded by temporal surface operators.
Hence, in this high temperature phase we have
\begin{equation}
\langle {\cal O}_{\text{temporal}} \rangle \; \ne \; 0
\label{TempHigh}
\end{equation}
and the spatial surface operators exhibit a ``volume law'':
\begin{equation}
\langle {\cal O}_{\text{spatial}} \rangle \; \sim \; e^{-\text{Volume}(D)}
\label{SpatHigh}
\end{equation}
since in the AdS black hole geometry the warp factor is bounded from below,
as illustrated in Figure~\ref{amoeba}.
Since finite temperature breaks supersymmetry explicitly and makes scalars and fermions massive,
this behavior is expected to be generic, in particular, present in pure gauge theory or more general $\CN=2$ gauge theories.

Also note, that in the limit $\beta \to 0$ the theory reduces to a pure (non-supersymmetric)
three-dimensional Yang-Mills theory, which is expected to exhibit confinement and a mass gap.
Since under this reduction a temporal surface operator turns into a line operator (supported on $\gamma$)
in the 3d gauge theory, the behavior \eqref{TempHigh} is certainly expected.


\bibliographystyle{JHEP_TD}
\bibliography{surface}

\end{document}